\documentclass[11pt,preprint]{aastex}
\usepackage{amsmath}
\usepackage{color}
\usepackage{graphicx}
\usepackage{natbib}
\begin{document}

\title{Colors of a Second Earth II: 
 Effects of Clouds on Photometric Characterization of Earth-like Exoplanets}

\author{
  Yuka Fujii\altaffilmark{1}, 
  Hajime Kawahara\altaffilmark{2},
  Yasushi Suto\altaffilmark{1,3,4}, 
  Satoru Fukuda\altaffilmark{5}, 
  Teruyuki Nakajima\altaffilmark{5},
  Timothy A. Livengood \altaffilmark{6}, and
  Edwin L.~Turner \altaffilmark{1,4,7}
}
\altaffiltext{1}{Department of Physics, The University of Tokyo, Tokyo 113-0033, Japan}
\altaffiltext{2}{Department of Physics, Tokyo Metropolitan University, Hachioji, Tokyo 192-0397, Japan}
\altaffiltext{3}{Research Center for the Early Universe, Graduate School of Sciences, The University of Tokyo, Tokyo 113-0033, Japan}
\altaffiltext{4}{Department of Astrophysical Sciences, Princeton University, Princeton, NJ 08544}
\altaffiltext{5}{Center of Climate System Research, The University of Tokyo, Kashiwa 277-8568, Japan}
\altaffiltext{6}{NASA/Goddard Space Flight Center
Greenbelt, MD 20771}
\altaffiltext{7}{Institute for the Physics and Mathematics of the Universe, The University of Tokyo, Kashiwa 277-8568, Japan}
\email{yuka.fujii@utap.phys.s.u-tokyo.ac.jp}

\begin{abstract}
As a test-bed for future investigations of directly imaged terrestrial exoplanets, we present the recovery of the surface components of the Earth from  multi-band diurnal light curves obtained with the EPOXI spacecraft. 
We find that the presence and longitudinal distribution of ocean, soil and vegetation are reasonably well reproduced by fitting the observed color variations with a simplified model composed of {\it a priori} known albedo spectra of ocean, soil, vegetation, snow and clouds. 
The effect of atmosphere, including clouds, on light scattered from surface components is modeled using a radiative transfer code. 
The required noise levels for future observations of exoplanets are also determined. 
Our model-dependent approach allows us to infer the presence of major elements of the planet (in the case of the Earth, clouds and ocean) with observations having S/N $\gtrsim 10$ in most cases and with high confidence if S/N $\gtrsim 20$. 
In addition, S/N $\gtrsim $ 100 enables us to detect the presence of components other than ocean and clouds in a fairly model-independent way. 
Degradation of our inversion procedure produced by cloud cover is also quantified. 
While cloud cover significantly dilutes the magnitude  of color variations compared to the cloudless case, the pattern of color changes remains. 
Therefore, the possibility of investigating surface features through light curve fitting remains even for exoplanets with cloud cover similar to the Earth's. 
\end{abstract}

\section{Introduction}

Exoplanet research is now a rapidly maturing branch of astronomy. 
In particular, a number of recently discovered planets have masses of order the Earth's and offer the possibility of studying other worlds much like our own. 
The initial analysis of the Kepler mission data indicates that the small transiting planets are even more abundant than large ones \citep{borucki2010, borucki2011} which is consistent with the statistical trend for planets discovered by the radial-velocity method \citep{howard2010}. 
Moreover, candidates of nearly Earth-sized exoplanets within the habitable zones of their host stars have been already claimed \citep[e.g.][]{vogt2010,borucki2011}. Nevertheless a mere detection of such a planet would be far from convincing evidence for habitability. 

Far richer information concerning the habitability would come from the atmospheric and surface properties of such planets, which will be probed by photometric and spectroscopic observation of planets in optical and near-IR bands. 
The light in the bands is dominated by starlight scattered by the planets, and the
wavelength-dependent reflectivity and absorption features could 
provide direct indications of the planetary surface and atmospheric properties. This is why a number of on-going and future efforts including SEEDS \citep{tamura2009}, TMT \citep[with a future  instrument for direct imaging of Earth-like planets, see][]{matsuo2010}, O3 \citep{savransky2010}, See-COAST \citep{schneider2009}, SPICA \citep[e.g.][]{enya2009} plan to develop instruments for direct observations of exoplanets. 

Until we have the first scattered light observation for an Earth-like exoplanet, we may use the Earth itself as a useful test-bed for future
investigations as proposed by \citet{ford2001}. 
One approach is to extract spatially integrated scattered light  from 
Earthshine. In addition to several clear absorption lines of
biologically important molecules like H$_2$O, O$_2$, O$_3$ and CH$_4$\citep[e.g.][]{desmarais2002}, a 
marginal signature of the red-edge due to vegetation \citep[sharp increase of reflectivity at wavelength longer than 0.75$\mu$m, e.g.][]{seager2005} has been reported
\citep[e.g.][]{woolf2002,arnold2002,rodriguez2006,hamdani2006} in an amount consistent with simulations \citep[e.g.][]{tinetti2006a, tinetti2006b, rodriguez2006}.

Another approach is based on remote-sensing observation of the Earth, notably multi-band photometric data obtained with the EPOXI spacecraft. \citet{cowan2009} performed principle component analysis (PCA) in order to reconstruct
surface features from the EPOXI data. Their PCA inversion 
identified two major eigen modes, the ``red'' and the ``blue''
components, which they roughly interpreted as land and ocean,
respectively. They attempted to map the red component on the
planetary surface using the diurnal variation, and found that it
roughly reproduces the land distribution of the Earth along the
longitudinal direction.  
We note here that their analysis neglects the
effect of clouds on the diurnal light-curve since they assumed that the
cloud cover fraction is constant at all time.  

\citet{oakley2009} also
attempted a longitudinal mapping of land and ocean components from a
single-band, instead of multi-band, simulated photometry using the difference in reflectivity between land and ocean.

\citet{fujii2010} (hereinafter Paper I) developed another reconstruction method of the planetary surface. In contrast to model-independent
approaches such as PCA, Paper I approximates the planetary surface by
a combination of four components: ``ocean,'' ``soil,''
``vegetation,'' and ``snow''. 
The fractional areas of the four components were fit to reproduce the simulated light-curves of multi-band photometry of the {\it cloudless} Earth, and the method recovers the presence of ocean, soil and even vegetation reasonably well. 

The present paper extends the methodology and analysis
of Paper I in various respects: 1) we now include a ``cloud'' component
both in computing simulated diurnal multi-band light-curves (forward
procedure) and in reconstructing the fractional areas of the five
components (inverse procedure), 2) the simulated light-curves are based on the radiative transfer code {\it rstar6b} \citep{nakajima1988}\footnote{OpenCLASTR: http://www.ccsr.u-tokyo.ac.jp/~clastr/} instead of employing 
Paper I's single scattering approximation, and 3) the reconstruction procedure
is applied to the EPOXI data instead of simulated light curves.

The rest of the paper is organized as follows.  Section \ref{s:EPOXI} describes the technical details  of the EPOXI observations on which the subsequent sections are based.  
Section \ref{s:sim} discusses our forward procedure for computing the simulated light curves using {\it rstar6b} with input of  
land albedo data obtained with the MODIS (MODerate resolution Imaging Spectroradiometer) onboard the Earth Observing Satellites {\it Terra} and {\it Aqua} plus cloud data from MODIS onboard the {\it Terra}. Our inverse procedure, {\it i.e.} the 
reconstruction of the fractional areas of different surface components,
is detailed in Section \ref{s:inv}. We also consider six different combinations of assumed soil, clouds and atmosphere properties, and discuss the resulting systematic uncertainty of the reconstruction.  
Section \ref{s:dis} uses this improved methodology to investigate the effect of clouds and the limits they impose on the interpretations of planetary light curves. 
Section \ref{s:conclusion} summarizes our main conclusions. 
Appendix \ref{ap:input} supplements the description of pre-process of the input data, 
and Appendix \ref{ap:aitoff} derives the longitudinal map reconstructed from EPOXI data via our inverse procedure.

\section{EPOXI observation of the scattered light from the Earth}
\label{s:EPOXI}

\begin{figure}[!h]
  \centerline{\includegraphics[width=140mm]{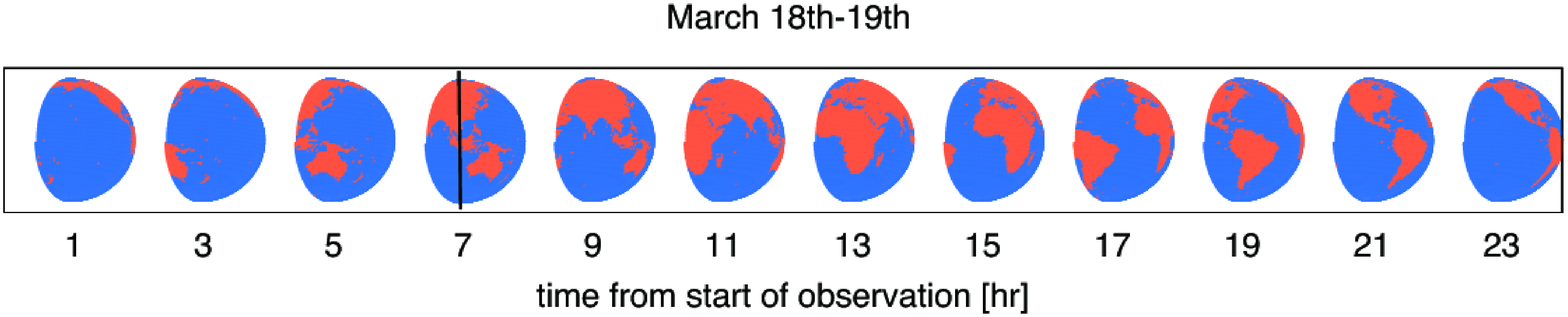}}
  \centerline{\includegraphics[width=140mm]{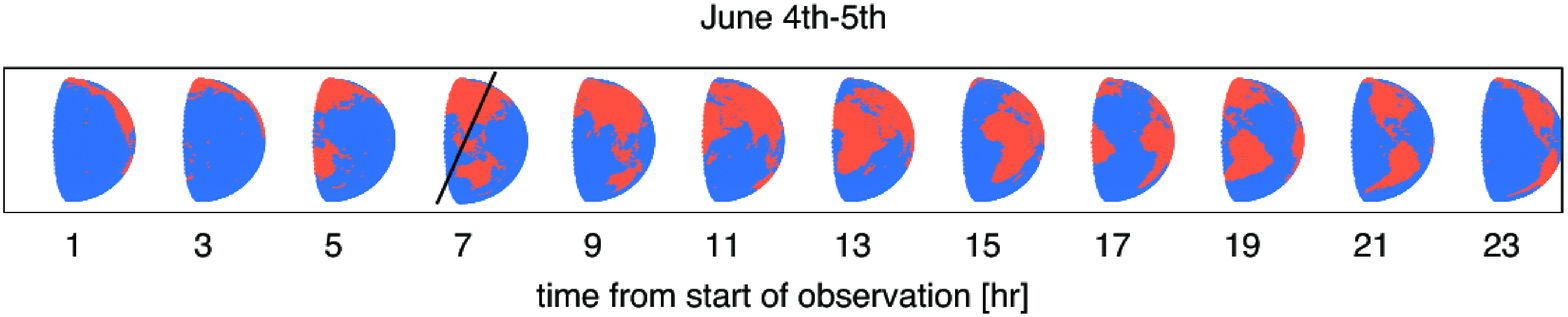}}
\caption{Snapshots of the observational geometry every two hours during EPOXI observations of the Earth on 18-19 March 2008 (top) and on 4-5 June 2008 (bottom). The black solid line at hour 7 indicates the orientation of the Earth's rotational axis on each day. }
\label{fig:snapshot}
\end{figure}

\begin{figure}[!htbp]
  \centerline{\includegraphics[width=100mm]{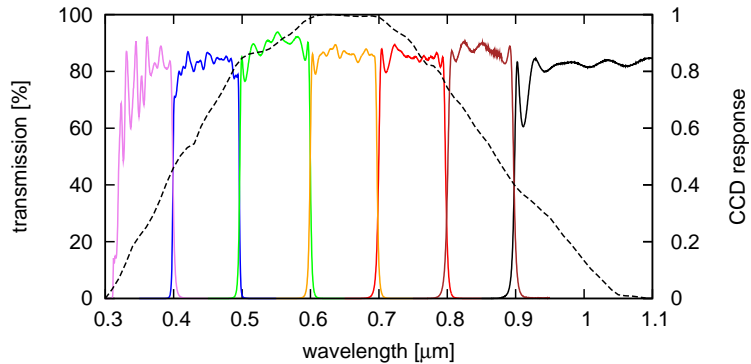}}
\caption{
Solid: filters of the high-resolution visible-light instrument (HRIVIS) on the Deep Impact spacecraft, used for the EPOXI mission. The 'violet' filter (nominally 0.35$\mu $m) is short pass, where the detector and optical system components limit short wavelength sensitivity to $\geq 0.365 \mu {\rm m}$. The `IR' filter (nominally 0.95$\mu $m) is long pass, which is cut off at $\sim 0.97\mu{\rm m}$ by CCD response and optical system components. Dashed: CCD response function. }
\label{fig:EPOXIfilter}
\end{figure}

\begin{figure}[!h]
\begin{minipage}{0.5\hsize}
\begin{center}
  \includegraphics[width=70mm]{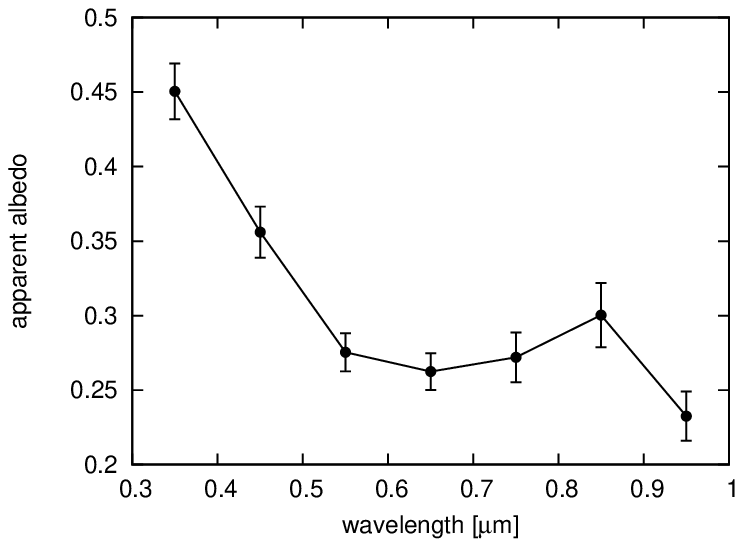}
\end{center}
\end{minipage}
\begin{minipage}{0.5\hsize}
\begin{center}
  \includegraphics[width=95mm]{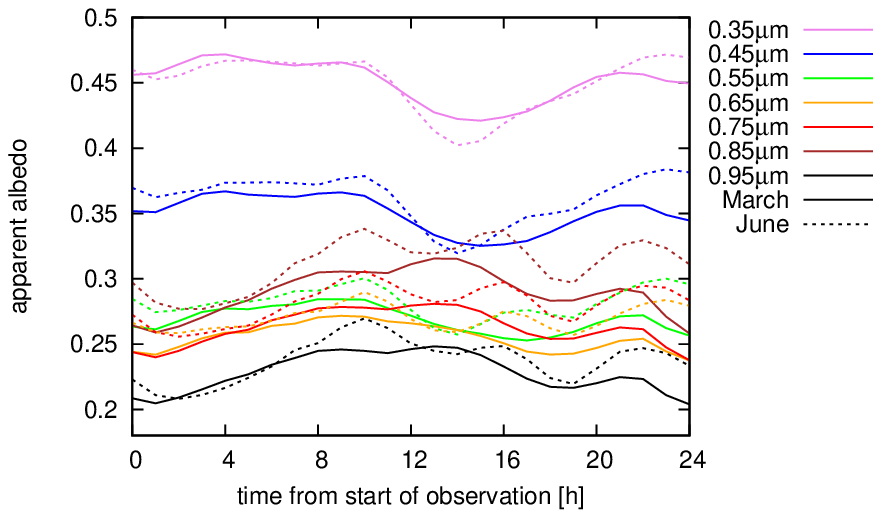}
\end{center}
\end{minipage}
\caption{Apparent albedo of scattered light from the Earth observed by EPOXI. 
Left: the photometry of 7 EPOXI bands averaged over the March and June data. Error bars show the standard deviation of the hourly flux measurements in each wavelength band and are dominated by actual variability, not measurement errors. 
Right: light curves of $0.37-0.4$ (purple line), $0.41-0.5$ (blue line), $0.5-0.6$ (light blue), $0.6-0.7$ (green), $0.7-0.8$ (red), $0.8-0.89$ (brown), and $0.9-0.97\,\mu {\rm m}$ (black) in the March data (solid) and June data (dashed). The observational uncertainty is less than 0.1 \%, which is negligible in this figure.}
\label{fig:lc_EPOXI}
\end{figure}

One significant improvement over Paper I of the present work is the use of direct observations of the Earth's diurnal light curves, both as a validation of our forward procedure and as input for the inverse procedure. 
In particular we employ observations of the Earth that were obtained during NASA's EPOXI mission. 
The EPOXI mission reused the Deep Impact spacecraft to carry out multi-band photometry of the Earth in optical/near-IR
\citep{livengood2011}.  
EPOXI observations of the Earth covered one whole day in each of March, May and June of 2008.  The May data include a lunar transit of the Earth and thus introduce  complexities that are beyond the scope of the 
present effort; we therefore do not use the May data. 

Figure \ref{fig:snapshot} depicts the geometry of the Earth during the March and June 2008 observing runs.
The spacecraft observed the Earth 76.7\% illuminated at a phase angle of 57.7$^{\circ}$ in March, near the 
northern spring equinox, and 61.5\% illuminated at a phase angle of 76.6$^{\circ}$ in June, near the northern summer solstice.  The data effectively mimic observations of Earth-twins near quadrature in exoplanetary systems because the distance between the spacecraft and the Earth in March 
($\sim 2.7 \times 10^7 {\rm km} = 0.18 {\rm AU}$) and in June ($\sim 5 \times 10^7{\rm km} = 0.34 {\rm AU}$) is a few thousand times the Earth's radius and thus represents a good approximation to the geometry of an observation from infinite range.  We use photometry obtained with the Deep Impact HRIVIS instrument at one-hour intervals in each of seven bands that span the visible range: $0.37-0.4\mu {\rm m}$,
$0.41-0.5\mu {\rm m}$, $0.5-0.6\mu {\rm m}$, $0.6-0.7\mu {\rm m}$, $0.7-0.8\mu {\rm m}$,
$0.8-0.89\mu {\rm m}$, and $0.9-0.97\mu {\rm m}$.  

Images obtained with the HRIVIS camera are processed automatically for bias-subtraction, flat-fielding, an absolute calibration \citep{klaasen2008}. Data acquired during the EPOXI mission \citep[e.g.][]{ballard2010} indicate that the pixel-by-pixel flat-fielding correction has become inaccurate by about 1\%, with random distribution across the detector array. The EPOXI Earth observations combine signal from thousands of pixels per frame, reducing the significance of random deviations to negligibility. 

Photometric signal is extracted from EPOXI Earth images by aperture photometry, summing over a circular extraction region centered on Earth's disc. Background signal is estimated from an annular region of similar surface area. The calibrated image units multiplied by the known solid angle per pixel yields collected intensity in units of W/m$^2$/$\mu $m. The intensity is corrected for changes in the range from Earth to spacecraft, which varies slightly over the course of each 24-hour observing sequence and varies significantly between observations. Variations in Earth's heliocentric distance, while minor, are corrected similarly.

Measurement uncertainties in the photometry are very small, of order 0.1\% \citep{livengood2011}, evaluated from the distribution of pixel values in the annular background used in the photometry. Photometry of the Moon, which did not alter its aspect during a set of EPOXI measurements in May 2008, yields an independent estimate of photometric uncertainty of about 0.05\% after scaling for the relative diameter of Earth and Moon. Actual variability of the Earth's signal is a much greater source of variability in the measurements, creating a range of measurements that is approximately $\pm $4-8\% of the mean signal measured in each filter. This range of variability is defined by the limits of the central 68th-percentile of the photometric sample population, yielding ``one sigma'' confidence limits on the expected value to be determined from a single independent measurement. 

The complete data set and data-reduction procedures are described by \citet{livengood2011}. The HRIVIS filter functions are shown in Figure \ref{fig:EPOXIfilter} along with the CCD response function. 
Optical components within the telescope and spectrograph further limit the extreme blue and red sensitivity of the shortest and longest wavelengths bands, respectively.

\begin{figure}[htbp]
 \centerline{\includegraphics[width=120mm]{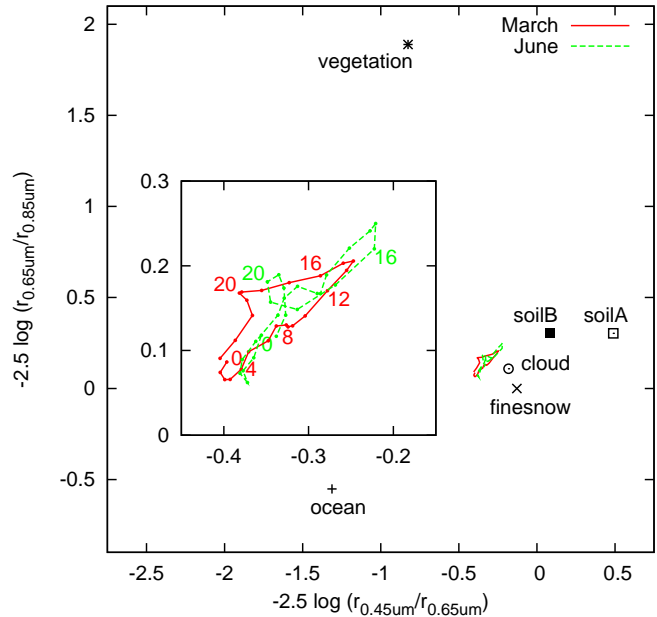}}
 \caption{Color-color diagram of EPOXI data from March (solid, red) and June  2008 (dashed, green). Numbers indicate hours from the start of the observation, as indicated in Figure \ref{fig:snapshot}. Points represent the colors of some representative
 Earth-surface components. }
  \label{fig:colorcolor_24-46_EPOXI}
\end{figure}

In Figure \ref{fig:lc_EPOXI}, we display the EPOXI data interpreted in terms of apparent albedo $p^*$. 
We follow the definition of $p^*$ in \citet{qiu2003}, as the ratio of total scattered intensity to the intensity of the incident solar flux scattered by a lossless Lambert sphere of the planet's radius, at the observed phase angle, 
\begin{eqnarray}
p^*(\lambda ) &=& \frac{I(\lambda )}{\frac{2}{3} F_* R_p^2 \Psi (\alpha )}, \\
\Psi (\alpha ) &=& \frac{(\pi - \alpha) \cos \alpha + \sin \alpha }{\pi}
\end{eqnarray}
where $I(\lambda )$ is the scattered intensity as a function of the wavelength $\lambda$, $F_*$ is the incident flux,  $R_p$ is the planetary radius, $\alpha $ is the phase angle, and $\Psi (\alpha )$ is the phase function for scattering by a Lambert sphere. Note that it is 3/2 times the geometric albedo. 

The left panel of Figure \ref{fig:lc_EPOXI} displays the photometry of 7 EPOXI bands averaged over both of the observed days, with the standard deviation of hourly measurements indicated as error bars. The variability in the measurements does not result from finite observational precision, but instead is the actual variability of the Earth's signal. 
As mentioned above, the observational uncertainty of the EPOXI measurements results in negligible noise level ($\le $ 0.1\% with no exposure time longer than 100 ms) because the source (the Earth) is extremely bright. 
The right panel of Figure \ref{fig:lc_EPOXI} shows the diurnal variation of each color band in
March (solid) and in June (dashed); this figure illustrates both the common patterns of and difference between the March and June data. 
These data are qualitatively consistent with the expected photometric variability of the Earth, as first discussed by \citet{ford2001}.

The shape variations between the lightcurves in different filters suggest looking more closely at the systematic diurnal color variability. 
We define color quantitatively as 
\begin{equation}
C_{ab} = -2.5 \log _{10} \left( \frac{ r_a}{r_b} \right), \label{eq:color} 
\end{equation}
where $r_x$ is the reflectivity in band $x$, following the conventional astronomical practice.  We select the $0.41-0.5$ and
$0.6-0.7\mu {\rm m}$ bands to measure the variation in the blue region of the spectrum at short wavelengths and the $0.6-0.7$ and $0.8-0.89 \mu {\rm m}$ bands to quantify the variation in the red region of the spectrum at long wavelengths.
Figure \ref{fig:colorcolor_24-46_EPOXI} displays the resulting color-color diagram of the diurnal light curves of the Earth for the two days. 
The colors of some representative components as seen through clear atmospheric layers are plotted as points in Figure \ref{fig:colorcolor_24-46_EPOXI} for reference. 

Unsurprisingly, the average colors of the Earth are similar to the colors of clouds, but somewhat displaced due to the surface components.

\section{Forward procedure: simulating the diurnal 
scattered light-curve of the Earth}
\label{s:sim}

Our goal is to develop a methodology to reconstruct the surface
properties from the diurnal scattered light curves of exoplanets in habitable zones. 
However, before proceeding to this inversion, we choose to gain better insights into the problem and to validate our assumptions and models by carrying out forward calculations and comparing the results to the EPOXI data. 
In particular, we aim to determine the accuracy and limitation of the models of surfaces, clouds, and atmospheres as well as to validate our simulation code. 

Therefore, in this section we combine the models and empirical data of the surface and the cloud distribution so as to reproduce the diurnal
light curves in multi-bands observed by EPOXI.  In addition to the forward calculation presented in Paper I, numerous other groups have simulated the scattered light of the Earth \citep{ford2001, tinetti2006a, tinetti2006b, rodriguez2006, palle2008, oakley2009, arnold2009, doughty2010,robinson2011}. 

Basically the computation of the light-curves of the Earth requires the
integration of the scattered light over the illuminated and visible surface. We have to divide the surface into two
dimensional pixels and sum up the contribution from each pixel by
approximating it locally as a flat plane.  We use
$2.0^{\circ}\times2.0^{\circ}$ latitude/longitude pixels as a compromise
between computation time and accuracy.

Unlike Paper I, which neglects multiple scattering and molecular 
absorptions, we incorporate both effects by using the line-by-line radiative 
transfer code {\it rstar6b}\footnote{http://www.ccsr.u-tokyo.ac.jp/~clastr/dl/rstar6b.html}.  This 
code has been developed intensively as a tool for interpreting the remote-sensing data of the Earth. It incorporates the detailed modeling of optical properties of the atmosphere and clouds.

\subsection{Model assumptions in the forward procedure}

\subsubsection{Atmosphere and clouds}

Scattering processes in the Earth's atmosphere are   complicated.  Since the albedos of clouds are generally much higher than those of the planetary
surface components, the broad distribution of the cloud optical depths 
crucially affects the light curve of the Earth.  Atmospheric pressure is
the key parameter that determines the optical depth of Rayleigh
scattering as well as the atmospheric features due to  molecules, and the spectral features depend sensitively on the composition of the atmosphere. Both affect the broad-band photometry considered here.

For simplicity we adopt a representative set of parameters called the ``US standard'' model atmosphere \footnote{U.S. Standard Atmosphere, 1976, U.S. Government Printing Office, Washington, D.C., 1976}, and assume that the model is valid everywhere. Basically the model provides the molecular composition and the temperature-pressure relation as a function of the altitude. However, the total scattered light is not sensitive to the details of those profiles. 

In addition to 
the homogeneous ``US standard'' model atmosphere, our model includes an empirical inhomogeneous cloud distribution.  
The spatial distribution of clouds is characterized by the cloud cover fraction $f$ and the cloud optical thickness $\tau _{\rm cld}$ in each pixel. 

We use the cloud data set from Terra/MODIS Atmosphere Level 3 Product\footnote{http://ladsweb.nascom.nasa.gov/} \citep{MODIS}. 
These data are daily global maps based on MODIS onboard the Earth Observing Satellite {\it Terra} and are given in $1.0^{\circ}\times1.0^{\circ}$ latitude/longitude pixels. 
Thus we merge them into $2.0^{\circ }\times 2.0^{\circ }$ pixels to match the resolution that we use in the numerical code. 
The cloud data are not available for some locations, and we interpolate the data according to the
procedure described in the Appendix \ref{ap:input} to fill these gaps. 

Although we take account of the spatial distribution of clouds, their optical properties are assumed to be everywhere the same. 
These optical properties depend on various parameters including the vertical profile, composition, phase (liquid or solid), radii and shapes of particles.  
For simplicity, we assume that all of the clouds are located at the altitude of 3.5-6.5 km and are composed of pure water; we neglect aerosols.  The size distribution function of cloud particles is assumed to be 
\begin{equation}
D(r; r_m) = \left\{
\begin{array}{ll}
 C \exp \left[ -\frac{(\ln(r/r_m))^2}{2(\ln(\sigma) )^2} \right] 
&(1.0\times 10^{-6}{\rm cm} \le r \le 3.0 \times 10^{-3}{\rm cm}) \\
 0  &({\rm otherwise})
 \end{array}
 \right.
\end{equation}
where $C$ is the normalization factor. 
It also uses the size parameter $r_m=8.0\times10^{-4}$cm and $\sigma =1.5$. 
Clouds with these properties are hereinafter called ``middle clouds''. 

\subsubsection{planetary surface components: land, ocean and snow}
\label{sss:surf}

The scattering model for the planetary surface is basically the same as
adopted in Paper I. 
However, there are small differences in our treatment of land Bidirectional Reflection Distribution Functions (BRDF). 
These are described below. 

The scattering properties of land are described by the Rossi-Li model, a
parametrized BRDF. While the model has three different terms (isotropic,
geometric, and volume terms; see Paper I for details), we now approximate the
planetary surface as Lambertian (isotropic scattering surface), and
neglect the geometric and volume terms. 
At the geometry of the EPOXI observation, this approximation leads to $\le $ 5\% differences in reflectivity compared to the light curves simulated with all three BRDF terms. 

We also follow Paper I in using the BRDF coefficients from ``snow-free gap-filled MODIS BRDF Model Parameters''\footnote{http://modis.gsfc.nasa.gov/}, which is a spatially and temporally averaged
product derived from the 0.05$^\circ$ resolution BRDF/albedo data. Therefore we again merge the data into
$2.0^{\circ}\times2.0^{\circ}$ pixels by simply averaging the BRDF
coefficients within each pixel (from $40\times40$ data points). 

The BRDF for ocean pixels is described in Appendix B of Paper I \citep[see also][]{nakajima1983}. 
The key parameter in this BRDF
is the wind velocity 10 m above the surface, which we set to 4 m/s everywhere.  
Note that this model neglects the sub-surface scattering. 
In reality, the sub-surface scattering results in a slightly bluer ocean color than our model \citep[e.g.][]{mclinden1997}. 

Finally we take account of the snow cover on the planetary surface, neglected in Paper I. 
The fraction of the snow-covered area is very small, but it may affect the
total scattered light curve because of its high albedo.  For this 
purpose, we use a data set of the monthly mean ice/snow cover fraction
from International Satellite Cloud Climatology Project (ISCCP)\footnote{http://isccp.giss.nasa.gov/index.html}.  The data are provided
in $2.5^{\circ}\times2.5^{\circ}$ grids, and we interpolate them into
$2.0^{\circ}\times2.0^{\circ}$ pixels according to the nearest-grid-point assignment.  We regard those pixels with snow cover
fraction greater than 0.5 as snow pixels, and then assign the albedo of
``fine snow'' from ASTER spectral library (shown in the left panel of
Figure \ref{fig:albedo_surface} below).

\subsection{Radiative transfer calculation with {\it rstar6b}}

\begin{table}
\caption{Grid system for radiance look-up table.}  \label{tab:grid}
\begin{center}
\begin{tabular}[h]{ccc} \hline \hline
parameter & symbol & grid \\ \hline
solar zenith angle	&	$\theta _0$	&	$5 + 10i \;\;\; (i=0, 1, ... 8)$\\
zenith angle of observation	& $\theta _1$	&	$5 + 10i \;\;\; (i=0, 1, ... 8)$\\
 azimuthal angle	&	$\phi$		&	$5 + 10i \;\;\; (i=0, 1, ... 17)$\\
 cloud optical thickness	&	$\tau _{\rm cld}$	&	$\tau _{{\rm cld,}\,0} = 0, \;\; \tau _{{\rm cld,}\,i} = 2^{i-1}	 \; (i=1, 2, ... 9)$ \\
 surface albedo & $a_{\rm surf}$ & $a_{{\rm surf,}\,i} = 0.05i \;\;\; (i=0, 1, ... 20)$ \\ \hline
 \end{tabular}
\end{center}
\end{table}

The scattering models for land, ocean and snow with atmosphere and clouds described above, allows us to run the
radiative transfer code {\it rstar6b}, and then sum up all of the 
contributions from the relevant pixels according to the assumed
geometry of the star-planet-observer system.

Apart from the surface BRDF model, the radiance of a pixel is determined by five  
parameters; the solar zenith angle ($\theta _0$) for the incident ray
(assumed to come in parallel), the zenith angle of the observation
($\theta _1$) for the scattered ray, the azimuthal angle between the incident ray and the observer($\phi $), the optical thickness ($\tau_{\rm cld}$) and the cloud cover fraction ($f$) (Table \ref{tab:grid}). 

Our surface classification proceeds as follows; first we identify pixels of ocean, and assign the corresponding BRDF model for ocean which 
includes strongly anisotropic specular reflection. 
Next we adopt an isotropic term only from the Rossi-Li BRDF for the 
remaining pixels (i.e., land) as described above.  The coefficient for 
the isotropic term is obtained from MODIS. 
Finally pixels covered by snow are identified according to the ISCCP data, then the albedo for those pixels is replaced by that of
``fine snow'' (shown in the left panel of Figure \ref{fig:albedo_surface} below). 

Then the total radiance of a pixel is given as the sum of the radiance of the surface through clear and cloudy portions of the atmosphere above that pixel:
\begin{equation}
  {\rm Rad}_{\rm total}(\theta_0, \theta_1, \phi, \tau_{\rm cld}, a) 
= (1-f)\,{\rm Rad}(\theta_0, \theta_1, \phi, 0, a) 
+ f\,{\rm Rad}(\theta_0, \theta_1, \phi, \tau_{\rm cld}, a).
\end{equation}
Therefore one has to compute ${\rm Rad}(\theta_0, \theta_1, \phi,
\tau_{\rm cld}, a)$ for each pixel by solving the radiative transfer
equation in principle.  
In practice, however, a significant gain in the 
computation time is achieved by tabulating ${\rm Rad}(\theta_0, \theta_1, \phi, \tau_{\rm cld}, a)$ and computing the radiance for each pixel via 
a linear interpolation of the resulting 5-dimensional look-up table. 
The grids for the five parameters are shown in Table \ref{tab:grid}. 
This procedure is repeated for each of the 7 EPOXI bands shown in Figure \ref{fig:EPOXIfilter}.

\subsection{Comparison of simulated light-curves with the EPOXI data}
\label{ss:modelvalidation}

In the preceding subsections, we have described our forward procedure to
compute the diurnal light curves of the Earth.  We now compare the resultant simulated curves with the data taken by EPOXI on March 18-19th and June 4-5th, 2008 so as to validate the procedure.

The geometric configuration of the Sun, the Earth and the observer is fixed for each EPOXI observing run.  
We use the cloud data from the Earth Observing Satellite {\it Terra}
for the corresponding days. 
The incident flux from the Sun to the Earth is approximated to be plane-parallel, and we adopt the
distant-observer approximation at the location of EPOXI ($\sim 5 \times 10^7{\rm km}$ away from the Earth).  In addition we can safely neglect
both the orbital motion of the Earth ($\sim 1^{\circ}$/day in the orbital phase angle), and the motion of EPOXI spacecraft during each observing run. 

\begin{table}
\caption{EPOXI photometric bands and the corresponding MODIS bands.} \label{tab:band_comp}
\begin{center}
\begin{tabular}{ccc} \hline \hline
  EPOXI band[$\mu$m] & central wavelength[$\mu$m] & MODIS band[$\mu$m] \\ \hline
  $0.3-0.4$ & 0.35&0.459-0.479 \\
  $0.4-0.5$ & 0.45&0.459-0.479 \\
  $0.5-0.6$ & 0.55&0.545-0.565 \\
  $0.6-0.7$ & 0.65&0.620-0.670 \\
  $0.7-0.8$ & 0.75&0.620-0.670 \\
  $0.8-0.9$ & 0.85&0.841-0.876 \\
  $0.9-1.0$ & 0.95&0.841-0.876 \\ \hline
\end{tabular}
\end{center}
\end{table}

Since our land BRDF coefficients are defined according to the MODIS
bands, they do not strictly match the EPOXI photometric bands.  We assign the MODIS data to the EPOXI bands as shown in Table 
\ref{tab:band_comp}. Three EPOXI bands ($0.3-0.4 \mu{\rm m}$,
$0.7-0.8\mu{\rm m}$, and $0.9-1.0 \mu{\rm m}$) do not have direct counterparts in the MODIS bands, which may affect the results to some
extent.

Then we simulate the {\it instantaneous} light curve at one hour intervals. Since the EPOXI exposure times are $\le$ $6.15\times10^{-2}$ sec, this is an excellent approximation. 

\begin{figure}[!h]
  \centerline{\includegraphics[width=80mm]{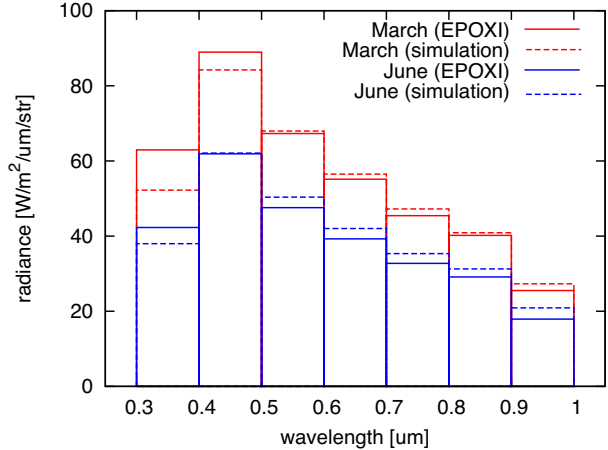}}
\caption{The 1-day average photometry of 7 EPOXI bands (solid) and those from simulated light curves (dashed). The March data is shown in red and June in blue. Note that the difference in radiance between the March and June data is due to difference in phase, i.e., the illuminated and visible area is bigger for March observation. }
\label{fig:sp_comp_Ty}
\end{figure}

\begin{figure}[!h]
\begin{minipage}{0.5\hsize}
\begin{center}
  \includegraphics[width=80mm]{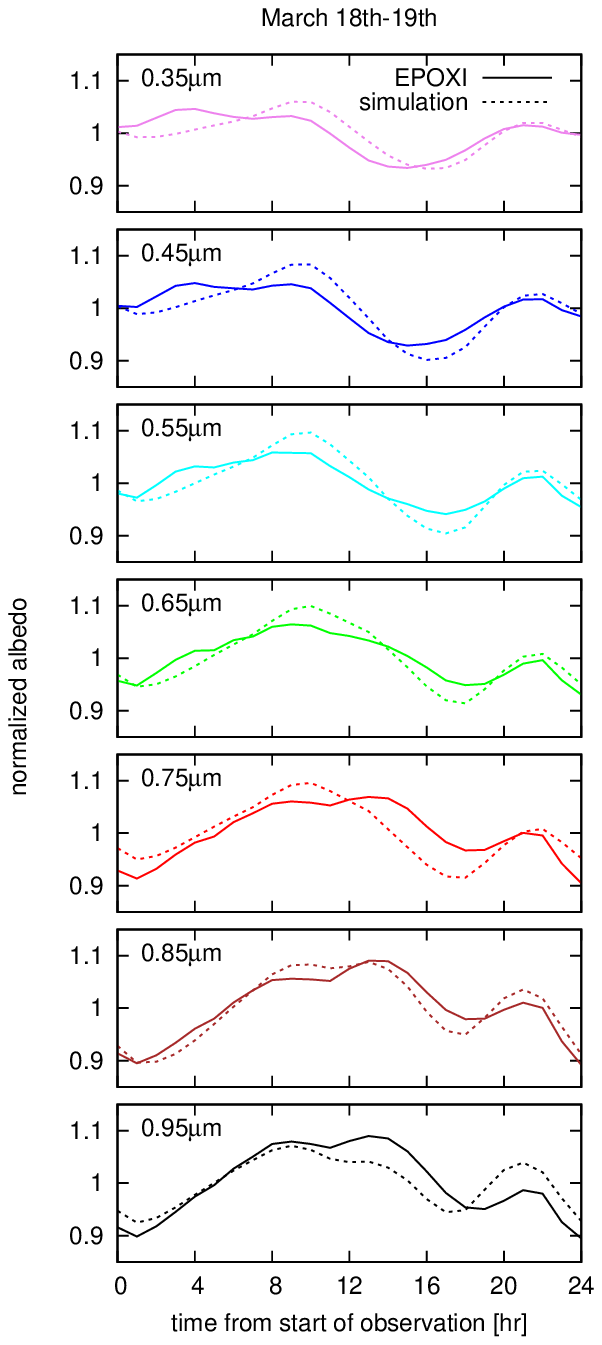}
\end{center}
\end{minipage}
\begin{minipage}{0.5\hsize}
\begin{center}
  \includegraphics[width=80mm]{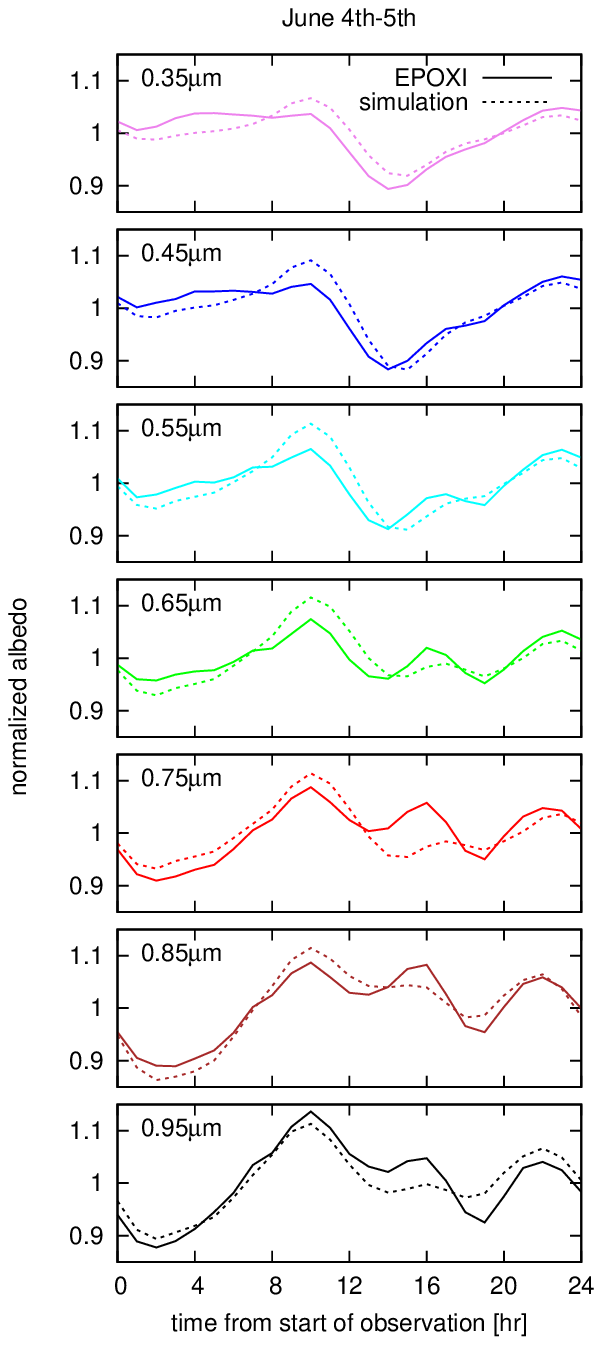}
\end{center}
\end{minipage}
\caption{Comparison of EPOXI (solid) and simulated (dashed) light curves, each normalized by its time average. Left: March. Right: June. }
\label{fig:lc_comp_Ty}
\end{figure}

Figure \ref{fig:sp_comp_Ty} compares the amplitudes of the total 
radiance between the EPOXI data (solid) and our simulations (dashed)
averaged over 24 hours; strictly speaking, we average 24 simulated
radiances sampled instantaneously at every hour in each band\footnote{Figures \ref{fig:sp_comp_Ty} and \ref{fig:lc_comp_Ty} are in the same format as Figures 3 and 4 of \citet{robinson2011}. }.
The agreement is within 10\% except for the $0.3-0.4\mu{\rm m}$ band.  
Figure \ref{fig:lc_comp_Ty} plots the normalized diurnal light curves, {\it i.e.,} the radiances each hour in each band divided by its 1-day average (plotted in Figure \ref{fig:sp_comp_Ty}); solid and dashed curves indicate the EPOXI and simulated 
data, respectively. The observed and simulated diurnal patterns are in good agreement, mostly
within 5\% accuracy. Even in the three bands ($0.3-0.4 \mu{\rm m}$, 
$0.7-0.8\mu{\rm m}$, and $0.9-1.0 \mu{\rm m}$) in which the EPOXI and
MODIS band correspondence is not so good, the difference is within 10\%.

Thus we conclude that our forward procedure reproduces the observation adequately for our purposes. 

In addition to the forward procedure adopted here, we have investigated several alternatives. 
In particular we simulated light curves by treating  all cloud pixels as (1) isotropic gray scatterers of albedo 0.4 which completely obscure the surface or  (2) $\tau _{\rm cld}=10$ clouds. 
Instead of {\it rstar6b}, we employed (1) an analytic 2-stream approximation \citep{meador1980} or (2) an alternative numerical code {\it libRadtran}\footnote{http://www.libradtran.org/}. 
Finally we calculated light curves (1) without account for the specular reflection of ocean water or (2) based on an alternative set of MODIS reflection spectra of surface components, called ``white sky'' albedo. 
None of these variations reproduced the EPOXI data more successfully than the adopted procedure and some gave considerably inferior performance.

\section{Inverse procedure: recovering the planetary surface}
\label{s:inv}

\subsection{Method}

So far we have developed a forward procedure to compute the multi-band diurnal light curves of the Earth, and confirmed that our model reproduces the EPOXI data reasonably well. Next we consider an inverse procedure to recover the Earth's surface components from EPOXI multi-band photometric data.
This exercise is intended to model the recovery of planetary surface information from future photometric data for terrestrial extrasolar planets. 
At this point, it is necessary to employ additional assumptions so that the inversion is possible.  In particular, we assume that the planetary surface consists of a finite number of scattering components each with {\it a priori} known isotropic reflection spectra. Five components (ocean, snow, vegetation, soil and cloud) are included and used to fit the diurnal EPOXI photometric data by a linear combination of their isotropically averaged albedo spectra.

Thus the apparent albedo $p_j^*(t_i)$ of the Earth at the $j$-th band at time $t_i$ is described as a linear combination of the $k$-th components:
\begin{eqnarray}
\label{eq:model} 
p_j^{*{\rm (model)}}(t_i) = D_{jk} A_k (t_i) , 
\end{eqnarray}
where $D_{jk}$ is the {\it effective albedo} of the $k$-th component at the
$j$-th band.  We define the {\it effective albedo} as the ratio of the
total scattered intensity of a sphere wholly covered with a single component and the cloudless atmosphere to that of a lossless ({\it i.e.}, non-absorbing) Lambert sphere {\it at the same phase}. The coefficient, $A_k (t_i)$, corresponds to the 
geometrically weighted area of the $k$-th components:
\begin{eqnarray}
A_k = \frac{1}{\pi} \int _{IV} 
m_k(\theta, \phi) \cos \vartheta _0(\theta , \phi ) 
\cos \vartheta _1(\theta , \phi ) d \Omega .
\end{eqnarray}
In the above expression, $\vartheta _0 (\theta , \phi )$ is the solar
zenith angle viewed from the planetary surface coordinate $(\theta,
\phi)$, $\vartheta _1 (\theta , \phi )$ is the zenith angle of the 
observer, $m_k (\theta, \phi)$ denotes the fractional area of the $k$-th
type at the coordinate $(\theta, \phi)$, and the integral is performed
over the illuminated (I) and visible (V) area of the Earth from the observer.

We should note here that equation (\ref{eq:model}) treats the cloud
cover simply as a single additional component to the other 4 surface components.  In reality, the clouds are very complicated and their scattering spectra are dependent on their thickness, altitude, composition, and particle size distribution. 
In addition, the scattering properties of clouds and those of the surface components below interact in a complex way. 
Thus a fully realistic model would be underdetermined and therefore is impractical. 
Instead, we choose to characterize the clouds  by a single representative reflection spectrum with a fixed set of properties and 
a single optical depth $\tau _{\rm cld}$ (our fiducial model is a middle
cloud with $\tau _{\rm cld}=10$; see Table \ref{tab:cldprop}).
The validity of this major simplification will be examined in section \ref{ss:dis2}.

Our inverse procedure is carried out by adjusting the values of $A_k(t_i)$ in equation  (\ref{eq:model}) to optimize the fit. 
In principle, the fit is performed by minimizing $\chi ^2$: 
\begin{equation}
\chi ^2 (t_i) = \sum _{j} 
\frac{\left[ p_j^{*{\rm (obs)}}(t_i) - p_j^{*{\rm (model)}}(t_i)\right]^2}
{\sigma _j^2 (t_i)}, \label{eq:chi2}
\end{equation}
with $\sigma _j (t_i)$ being the appropriate observational errors.  
However, the errors in the EPOXI data are very small, much smaller than those likely to be obtained in any future observations of exoplanets.
Thus we employ the following $\bar
\chi ^2$ as a measure of the goodness of the fit: 
\begin{equation}
\bar \chi ^2 (t_i) = \sum _{j} 
\left[ p_j^{*{\rm (obs)}}(t_i) - p_j^{*{\rm (model)}} (t_i)\right]^2. 
\label{eq:barchi2}
\end{equation}
The magnitude of $\bar \chi ^2 (t_i)$ implies the required observational photometric accuracy to achieve such a recovery of the surface properties (see Figure \ref{fig:barchi}). 

We imposed the condition $A_k(t_i) \ge 0$ for all $k$ and $t_i$ in order to avoid unphysical models. 
We minimized $\bar \chi^2 $ in equation (\ref{eq:barchi2}) using the Bounded Variable Least-Squares Solver (BVLS) developed by \citet{lawson1974,lawson1995}\footnote{The original code of the BVLS is available through NETLIB (http://www.netlib.org/lawson-hanson/index.html).}. 
The BVLS is designed to solve linear least square problems with bounded conditions on  variables.

\subsection{Albedo models}
\label{ss:albedomodels}

\begin{figure}[!htbp]
\begin{minipage}{0.5\hsize}
\centerline{\includegraphics[width=80mm]{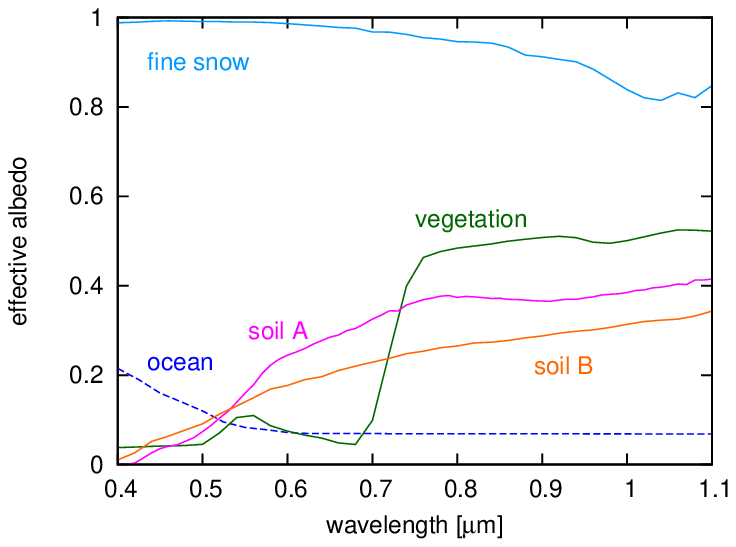}}
\end{minipage}
\begin{minipage}{0.5\hsize}
\centerline{\includegraphics[width=80mm]
{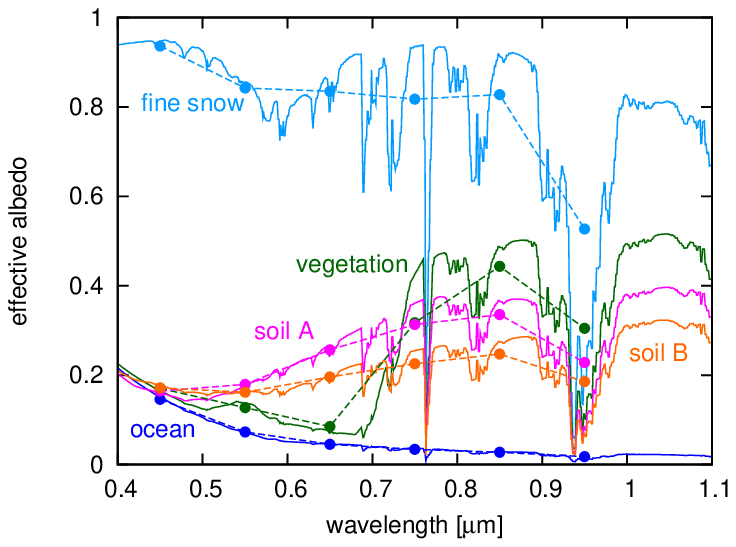}}
\end{minipage}
\caption{ The effective albedo spectra for fine snow, vegetation, soil A (``class: inceptisol''), soil B (``class: entisol'') and ocean. Left: Spectra from the ASTER library without atmosphere (solid lines).  The reflectivity of oceans  (including the scattering beneath the surface) is plotted as a dashed line for reference (courtesy of G. Tinetti; see also \citet{mclinden1997}). 
Right: Spectra with the US standard atmosphere (but neglecting the effect of cloud) at the phase of June 4th, 2008 (phase angle $\sim 76.6^{\circ}$) computed with {\it rstar6b}. Filled circles indicate the averages over the seven EPOXI photometric bands.
\label{fig:albedo_surface}}
\end{figure}

\begin{figure}[!h]
\centerline{\includegraphics[width=80mm]
{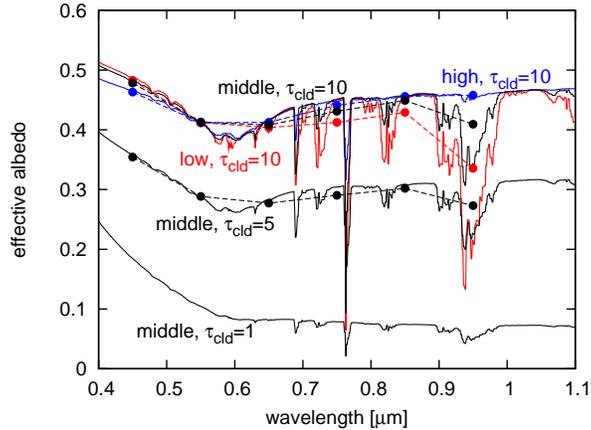}}
 \caption{Effective albedo of different cloud types at the June phase ($76.6^{\circ}$) calculated by {\it rstar6b}. Red: low clouds of
 optical thickness 10, blue: high clouds of optical thickness 10, and  black: middle cloud of optical thickness 1 , 5 and 10. The definitions for ``low'', ``middle''  and ``high'' clouds are given in Table \ref{tab:cldprop}. } \label{fig:effalbd_cld}
\end{figure}

\begin{table}
\caption{Models adopted in the inversion procedure.} 
\label{tab:invmodel}
\begin{center}
\begin{tabular}[h]{ccccc} \hline \hline
  model & soil & cloud altitude & cloud optical thickness & atmosphere
 \\ \hline  
  I & A & middle & 10 & US standard\\
  II & B & middle & 10 & US standard\\
  III & A & low & 10 & US standard\\
  IV & A & high & 10 & US standard\\
  V & A & middle & 5 & US standard\\
  VI & A & middle & 10 & no ${\rm O}_{\rm 3}$ \\\hline
\end{tabular}
\end{center}
\end{table}

\begin{table}
\caption{Assumed properties of three classes of clouds.}  
\label{tab:cldprop}
\begin{center}
\begin{tabular}[h]{cccc} \hline \hline
  & low cloud & middle cloud & high cloud \\ \hline
  altitude [km] & $\sim$3.5  & 3.5-6.5  & 6.5-20.5 \\
  pressure [mbar] & 680-1000 & 440-680 & 40-440\\
  component & pure water & pure water & pure ice \\ 
  size parameter $r_m$ [cm] & $8.4\times10^{-4}$ & $8.0\times10^{-4}$ & $2.0\times 10^{-3}$ \\ \hline
\end{tabular}
\end{center}
\end{table}

The design matrix $D_{jk}$ which corresponds to the effective albedo in the $j$-th band of
the $k$-th component is the essence of our model. 
The initial model employs the five components: ``ocean,'' ``snow,''
``vegetation,'' ``soil,'' and ``clouds.''  The corresponding spectra are 
computed with the radiative transfer code {\it rstar6b} assuming the US
standard atmosphere, which takes account of the
effects of molecular absorption and the atmosphere's pressure-temperature profile.

The left panel of Figure \ref{fig:albedo_surface} plots the input spectra of the surface albedo (without atmosphere) for snow, vegetation and 
soil taken from the ASTER spectral library.  We consider two representative models for soil (A and B) to investigate the effects of the variation in soil composition on the results.  More specifically, we
adopt ``fine snow'' for snow, ``grass'' for vegetation, ``Dark yellowish
brown micaceous loam (class: inceptisol)'' for soil A, and ``Brown to
dark brown sand (class: entisol)'' for soil B, where the nomenclatures quoted are those of the ASTER spectral library.  The ``soil'' assumed in Paper I corresponds to soil B in this paper. 
The scattering model for ocean surface is described in Section \ref{sss:surf}. 
The right panel of Figure \ref{fig:albedo_surface} is
the effective albedo spectra including the effects of atmosphere given by {\it rstar6b} at the phase of the June observing run.  Filled circles indicate the averages over the seven EPOXI photometric bands for
the incident solar spectra; these are the elements of the design matrix $D_{jk}$.

The effective albedos for clouds are computed by considering cloud layers above a zero-albedo surface. 
This assumption is valid for large cloud optical depth $\tau_{\rm cld}$, but underestimates the real albedo for smaller $\tau_{\rm cld}$. Nevertheless this is a practical necessity to make the inversion feasible. Figure \ref{fig:effalbd_cld} shows the reflection spectra of different cloud optical thicknesses and altitudes. We define
``low, '' ``middle,'' and ``high'' clouds as summarized in Table \ref{tab:cldprop}; the other parameters for the clouds are the same as
adopted in the forward procedure.

Figure \ref{fig:effalbd_cld} indicates that the cloud albedos are fairly insensitive to the wavelength. The cloud altitude primarily changes the strength of the absorption lines, and the cloud optical thickness changes the overall amplitude of the albedo, though not strictly in a linear fashion. 
Again, the filled circles indicate the elements of the design matrix $D_{jk}$.

Table \ref{tab:invmodel} lists six models corresponding to different combinations of cloud and soil properties. 
First we consider our fiducial model (model I)
in \S \ref{ss:modelI}, and then compare the results to the other models in \S \ref{ss:modeldepend}.

\subsection{Decomposing the EPOXI data with model I}
\label{ss:modelI}

\begin{figure}[!h]
\begin{minipage}{0.5\hsize}
\begin{center}
  \includegraphics[width=80mm]{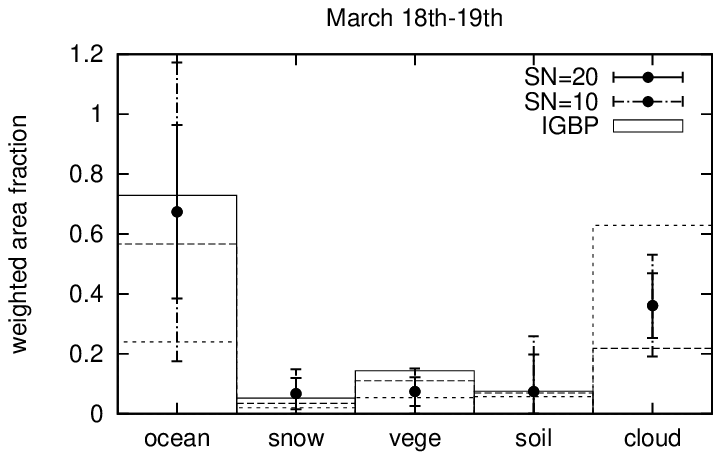}
\end{center}
\end{minipage}
\begin{minipage}{0.5\hsize}
\begin{center}
  \includegraphics[width=80mm]{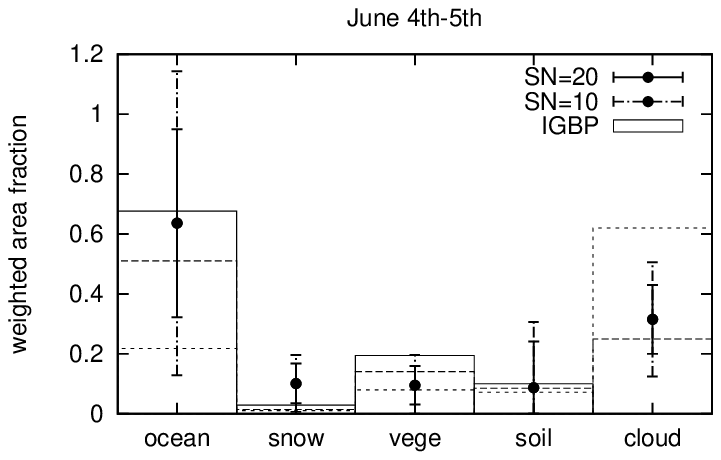}
\end{center}
\end{minipage}
\caption{Estimated fractional areas of the five surface components from the 1-day averaged data on March 18-19th, 2008 (left) and on June 4-5th, 2008 (right) for Model I.  
Filled circles and error bars are computed from 1000 Monte-Carlo realizations with S/N=20 (solid) and S/N=10 (dot-dashed). Dashed lines indicate reference values based on the
combination of IGBP classification, ISCCP ice/snow cover data and MODIS cloud cover on each observing day. These are intended to represent the ground truth. 
The histogram shown by solid lines displays the cloud-free case. 
The long- and short-dashed histograms exclude surface pixels with $\tau _{\rm cld} \ge 10$ and $\tau _{\rm cld} \ge 1$ clouds.} 
\label{fig:BVLS_June_cldml-10_soil-inceptisol_ave}
\end{figure}

\begin{figure}[!h]
\begin{minipage}{0.5\hsize}
\begin{center}
  \includegraphics[width=80mm]{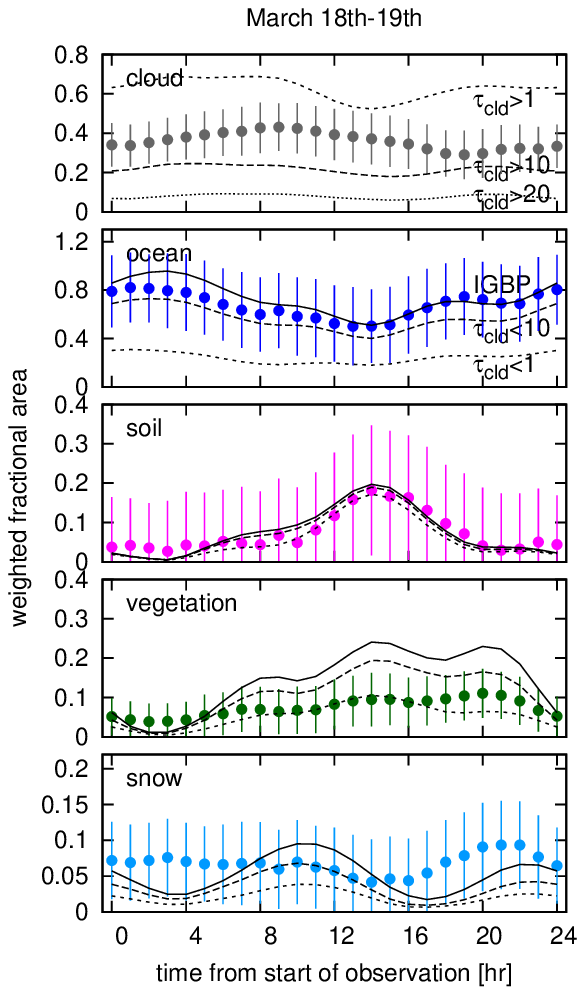}
\end{center}
\end{minipage}
\begin{minipage}{0.5\hsize}
\begin{center}
\includegraphics[width=80mm]{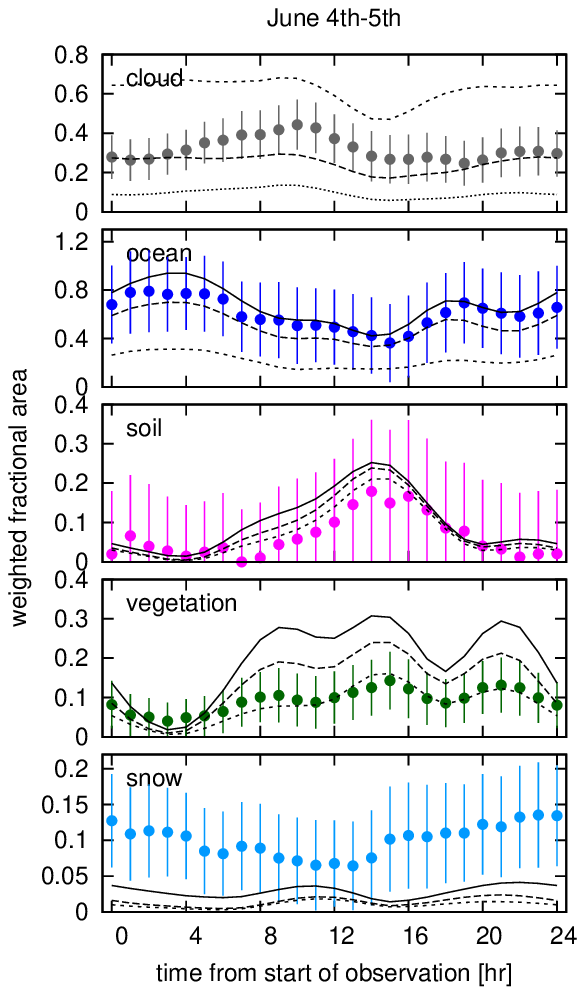}
\end{center}
\end{minipage}
\caption{The same as Figure
 \ref{fig:BVLS_June_cldml-10_soil-inceptisol_ave} but for the diurnal light curve data.  The filled circles and the errors are computed from
 1000 Monte-Carlo realizations with S/N=20 at every data point ({\it i.e.} each band and each time).}
 \label{fig:BVLS_June_cldml-10_soil-inceptisol}
\end{figure}

We apply our fiducial model (model I in Table \ref{tab:invmodel}) to the multi-band photometric data of the Earth observed by EPOXI, and infer
the fractional areas of the five components.  The results are plotted in Figure \ref{fig:BVLS_June_cldml-10_soil-inceptisol_ave} for
the 1-day average data, and in Figure
\ref{fig:BVLS_June_cldml-10_soil-inceptisol} at 1-hour intervals over the diurnal cycle, respectively. In both figures, the left panel shows the result for the observation on March 18-19th, 2008, while the right panel
on June 4-5th, 2008. 

The filled circles in both figures indicate the resulting weighted area
fraction $A_k$. The solid reference lines for ``ocean,'' ``soil,''
``vegetation,'' and ``snow'' are computed from the IGBP classification
map and ISCCP snow/ice cover data; we merge the original IGBP
classification with 17 classes into our 4 surface types by the same scheme 
as Table 2 of Paper I. Thus they correspond to the {\it cloudless}
Earth.  
More relevant estimates are given by the dashed reference lines in which the surface pixels covered by the clouds 
($\tau _{\rm cld}\ge 10$ for long dashed, and $\tau _{\rm cld}\ge 1$ for
short dashed) are removed from those components and regarded as cloud pixels (the case with $\tau _{\rm cld}\ge 20$ is also plotted in dotted lines in the ``cloud'' panels of Figure \ref{fig:BVLS_June_cldml-10_soil-inceptisol}).  The differences in 
reference lines of the left and right panels arise from the different
phase of the illuminated and visible part of the Earth (Fig \ref{fig:snapshot}), different cloud cover, and seasonal snow cover on the two different observing runs.

This inverse procedure seems to work {\it reasonably} well. In particular, ``ocean'' and ``clouds'' are recovered as dominant compared to the other components, and the estimated fraction of each component is slightly but systematically smaller than the IGBP values (solid lines).

Moreover, the diurnal variation pattern in Figure
\ref{fig:BVLS_June_cldml-10_soil-inceptisol} qualitatively follows
that of the actual surface components. For instance, the peaks of ocean at
$t$=4 and $t$=19 correspond to the Pacific and Atlantic oceans,
respectively.  The peak of soil corresponds to the Sahara desert, and
three peaks of vegetation to the Asian, African, and American continents.

The variation pattern of clouds is complicated by changes in the distribution of cloud optical thickness with time. 
For example, the inversion reproduces the $\tau _{\rm cld} \ge 1$ cloud curves best in amplitude, but matches $\tau _{\rm cld} \ge 20$ cloud variation better in shape. 

We note in passing that the time variation pattern of the coefficient of the ``blue'' eigen spectrum extracted by PCA, interpreted by \citet{cowan2009} to roughly correspond to ocean, does not reproduce the global distribution satisfactorily. 
They hypothesized that this results from the time variation of clouds, which they assume to be time independent.  Instead, the analysis presented here separates the ocean and clouds {\it a priori} and takes account of the time variation of the cloud cover. This is probably why our inversion reproduces the variation pattern of ocean more successfully. This implies that the blue component extracted from the model-independent PCA by 
\citet{cowan2009} does not correspond to ocean alone, but to a combination of ocean and clouds.

The time variation of each component can be translated into its longitudinal distribution \citep{cowan2008}.  By the
proper combination of the orbital motion of the planets, it would be
even possible to obtain a 2-dimensional map as first proposed by \citet{kawahara2010}.  In Appendix \ref{ap:aitoff} we exhibit the
longitudinal distribution of each component  obtained from the diurnal variation patterns.

\subsection{Signal-to-Noise requirement}

In order to estimate the feasibility of the inversion procedure given
the large observational errors anticipated in future observations of
Earth-like exoplanets, we create mock data sets with signal-to-noise ratios
of 20 and 10 using the EPOXI data. 
The variance of the estimated values is computed from 1000 Monte-Carlo realizations of the input spectra and is plotted as error bars in Figure 
\ref{fig:BVLS_June_cldml-10_soil-inceptisol_ave}.  Similarly the plotted
error bars in Figure \ref{fig:BVLS_June_cldml-10_soil-inceptisol} assume S/N=20. 

The errors in the estimated cloud cover are fairly small, while those of ocean fraction are significantly larger. 
This is simply due to the difference in their reflectivity; the lower the reflectivity is, the less accurate the inversion becomes. 
The errors of vegetation are relatively small, probably due to the strong signature of the red edge. 
At least in the case of the Earth, the presence of clouds and ocean is 
successfully inferred from the time-averaged data with S/N $\sim$ 10. 
However, their detectability is highly variable as the planet rotates and different portions of its surface are visible. 
So if the area of ocean on an extrasolar planet is smaller, its detection is therefore more difficult. For instance, a convincing detection of the presence of ocean with fractional area fraction of 
$\sim 0.4$ (roughly corresponding to the Earth at $t\sim 14$ in the March and June observing run) would require S/N more than 20. 

We may want to estimate the minimum requirement for direct imaging instruments to attain this S/N. 
In an extremely idealized case where the observational uncertainty comes from the photon noise of planetary light only, 
the S/N is approximately proportional to $\sqrt{N_{\rm photon}}$ where photon counts $N_{\rm photon} $ is scaled with equation (16) of \citet{fujii2010}. 
In this case, a 4 m telescope will be able to observe an Earth-twin at 10 pc away with S/N $\sim 20$ in 1 hour. 
The same S/N is also available with a smaller telescope if we carry out  hourly exposures for several days and phase them according to the planetary spin rotation period once it is determined \citep{palle2008}.

The photon counts will, however, substantially decrease through realistic future direct imaging instruments, due to the throughput and the quantum efficiency of CCD. 
Moreover, the dominant noise likely comes from speckles, exo-zodiacal light, and other instrumental noises such as readout noise and dark current. 
To be somewhat realistic, we consider the effect of throughput, quantum efficiency, readout noise, dark current, and exo-zodiacal light. 
The assumed instrumental design is listed in Table \ref{tab:conf2}, which is basically the same as Table 1 of \citet{kawahara2010}. 
On the assumption of these properties, S/N $\sim $ 20 roughly corresponds to 4 m telescope observations of an Earth-twin at 10 pc away with 1-hour exposure with the data stacked for 6 days and phased according to the rotation period (evaluated at 0.8-0.9 $\mu $m). 
The speckle noise depends on the details of the system which occults the light from the host star (coronagraph or external occulter). 
Since many such designs are under development, we do not try to account for noise due to residual light from the host star at this stage. 
Instead, the S/N requirement derived here will provide a goal of instrumental designs. 

\begin{table}[!tbh]
\begin{center}
\caption{Assumed observational parameters \label{tab:conf2}}
  \begin{tabular}{cccc}
   \hline\hline
Symbol	&	Quantity & Value & Unit	\\ \hline 
$\Psi $	&	sharpness	&	0.0433	&		\\
$\alpha	$	&	pixel scale	&	0.03125 &	arcsec/pixel \\
$h$	&	end-to-end efficiency &	0.5	&  \\
$\upsilon$	&	dark rate 	& 0.001	&	counts/sec	\\
$\kappa$		&	read noise 	& 2 		&	$\sqrt{{\rm counts}}$/read	\\
QE	&	quantum efficiency	 & 0.91	&	\\
$\Omega _z$	&	zodiacal light in magnitude &		23	&	mag/square arcsec	\\
$Z$	&	zodiacal light  &	1	&	\\
\hline
\end{tabular}
\end{center}
\end{table}

\subsection{Inversion Sensitivity to the model assumptions}
\label{ss:modeldepend}

\begin{figure}[!h]
  \centerline{\includegraphics[width=120mm]{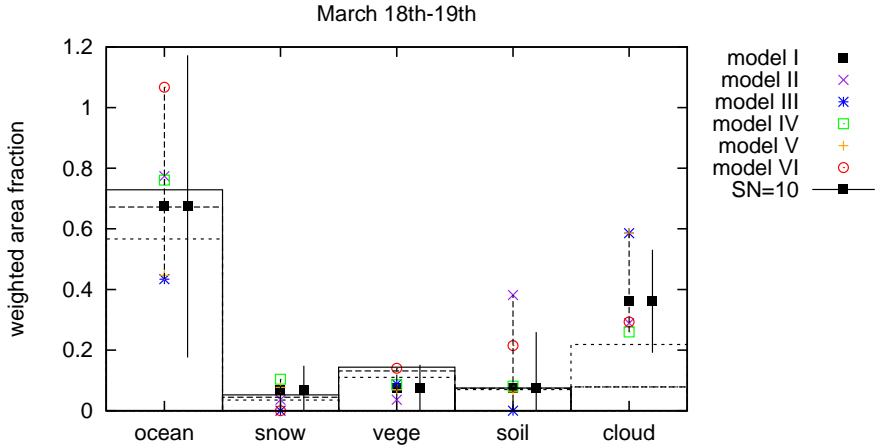}}
\caption{Decomposition of the averaged 7-band photometry of March with models I, II, III, IV, V and VI. 
The description of these models is given in Table
\ref{tab:invmodel}.  Histograms are the estimates of the ground truth identical to those in Figure \ref{fig:BVLS_June_cldml-10_soil-inceptisol_ave}.   
Vertical dashed lines 
connect the maximum and minimum values returned by the various models. The statistical variations produced by signal-to-noise ratio 10 are shown as  vertical solid lines. } 
\label{fig:wiperange_ave}
\end{figure}

\begin{figure}[!h]
\begin{minipage}{0.5\hsize}
\begin{center}
  \includegraphics[width=80mm]{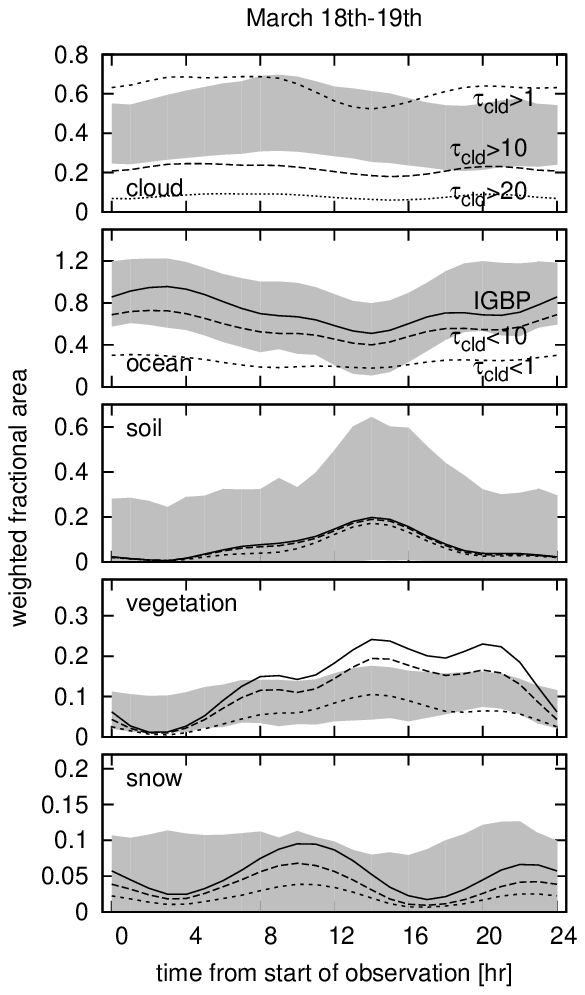}
\end{center}
\end{minipage}
\begin{minipage}{0.5\hsize}
\begin{center}
  \includegraphics[width=80mm]{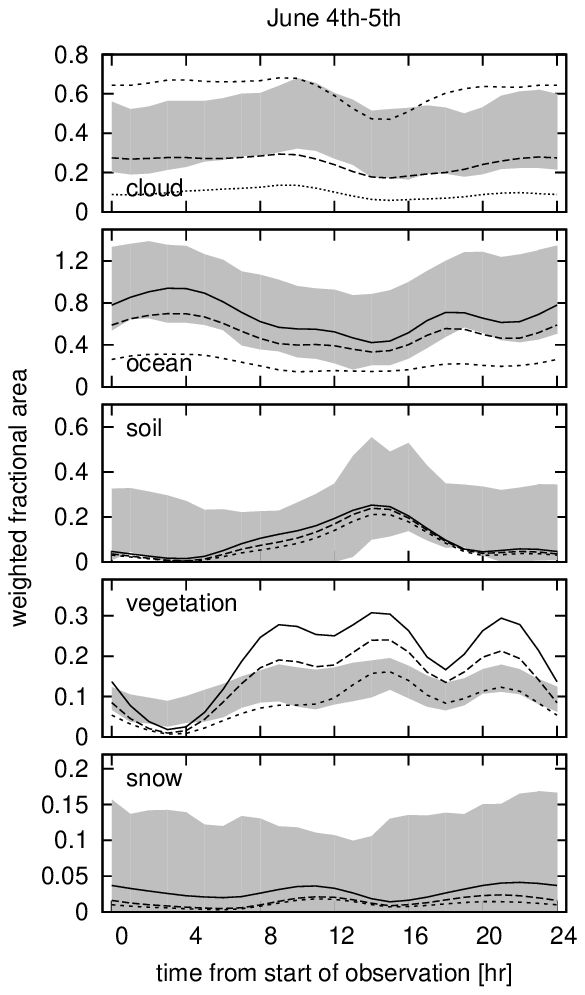}
\end{center}
\end{minipage}
\caption{The range of decompositions of the June light curves produced by models I, II, III, IV, V  and VI.  The shadowed region is
  bounded by the maximum and minimum values determined by this set of models.  } \label{fig:wiperange}
\end{figure}

As described earlier, our inversion may depend  heavily on the
adopted albedos for the selected surface components. It will not be easy to evaluate such systematic uncertainties associated with the model assumptions for future studies of exoplanets. 
However, we may gain some insights into the problem by comparing inversions of the EPOXI data using the different models listed in Table \ref{tab:invmodel}.  

Figure \ref{fig:wiperange_ave} is the same as the left  panel of Figure \ref{fig:BVLS_June_cldml-10_soil-inceptisol_ave} but for models I to
VI. Now the symbols indicate the estimates returned by the different 
models, and the vertical dashed lines indicate the range of model-to-model variations. 

Most of the model dependence can be qualitatively 
understood. 
Model II (cross), in which soil B is assumed instead of soil A, overestimates the area for soil because the effective albedo of soil B shows a lower-amplitude red portion of its spectrum (Fig \ref{fig:albedo_surface}). 
Models III (asterisk) and IV (open square) assume different altitudes for cloud layer. 
The difference in cloud altitude changes the absorption depths (Fig \ref{fig:effalbd_cld}), and therefore the estimated areas of all the components are affected accordingly. 
If we adopt model V (plus) with $\tau _{\rm cld} =5$ clouds, which have a reflectivity smaller than those with $\tau _{\rm cld}=10$, we end up with a larger estimate of the cloud area. 
Again, the estimation of all of the surface  components is also affected by this change. 
Model VI (open circle) assumes no ozone in the atmosphere, and the lack of ozone's broad absorption at $\lambda \sim 0.55\mu {\rm m}$ decreases the slope of all albedo models at short wavelengths. As a result, the area of ocean is estimated to be larger to compensate it. 

For reference we also plot the model I result with S/N=10 in Figure \ref{fig:wiperange_ave}.  The corresponding error bars for the fractional area are similar to the variation among different models considered here.  This implies that the model-dependent systematic errors are not dominant for an observational S/N of order 10. 

This conclusion is also consistent with the result for the diurnal light curves shown in Figure \ref{fig:wiperange}. The shaded region in each panel of this Figure indicates the range of the estimates among different models.  Again the basic features are similar to those shown in Figure \ref{fig:BVLS_June_cldml-10_soil-inceptisol}; while the amplitudes of the estimated area vary fairly significantly from model to
model, the overall variation patterns are relatively stable. Two of the  minor surface components, ``soil'' and ``snow'', are difficult to identify robustly, but the major components, ``cloud'' and ``ocean'',
are detectable in all the models.  Finally the presence of ``vegetation'' may be detectable at some phases without being confused
with the other components, probably due to its  conspicuous ``red edge'' feature.

\begin{figure}[!h]
  \centerline{\includegraphics[width=100mm]{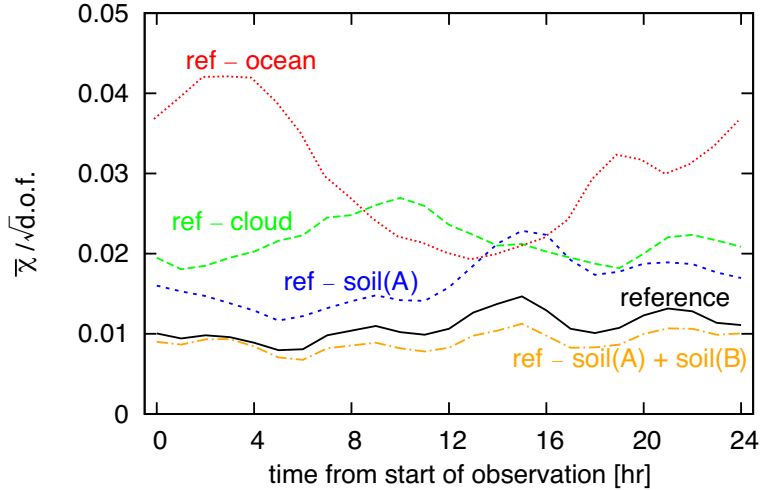}}
  \caption{The $\bar \chi $ of different models as a function of time. The curve labeled ``reference'' assumes ocean, soil A and middle clouds of optical thickness 10 but no vegetation component.  The curve labeled ``ref$-$clouds'' means that the model omits clouds but otherwise is identical to the reference model, and the other labels are to be understood in the same manner. }
  \label{fig:barchi}
\end{figure}

So far we have shown that our inversion procedure can identify ocean and
clouds reasonably well but that the other components are not securely identified. Nevertheless, we may ask if an additional, unknown surface component is required to achieve a satisfactory fit. 
This is a retreat in a sense but may be a less model-dependent approach. 

For this purpose, we compute and compare the goodness-of-fit to the observed data with and without specific surface types.  Figure
\ref{fig:barchi} shows $\bar \chi (t_i)$ per square-root of the degree-of-freedom for various models (without vegetation for
simplicity).  The reference model plotted in black solid line assumes that the planetary surface comprises of clouds, ocean and soil A.  The other curves correspond to the four variants of the reference model: models
with clouds, ocean and soil B instead of soil A (dot-dashed), with cloud and soil A without ocean (dotted), with ocean and soil A without clouds (long-dotted), and with clouds and ocean without soil (short-dotted). 

The resulting residuals of $\bar \chi/\sqrt{\rm d.o.f}$ may be roughly
interpreted as the inverse of the signal-to-noise ratio required to detect the discrepancy between the model and the observation. 
Since Figure \ref{fig:barchi} indicates that models with both ocean and clouds have values as low as 0.01, the presence of an additional
component would be revealed by the data only with S/N $\gtrsim $ 100. 
On the other hand, models without either clouds or ocean show the difference in $\bar \chi/\sqrt{\rm d.o.f}$ values of 0.02--0.03 at certain phases of the light curves. 
It provides a rough idea of the signal-to-noise ratio (or the amount of the exposure time) required to identify such surface components in future exoplanet observations.

Incidentally it is interesting to note that the time variation pattern of $\bar \chi/\sqrt{\rm d.o.f}$ essentially follows that of the {\it missing} component; for instance, the model without ocean shows the largest discrepancy at $t \sim 2$ when almost all of the illuminated and visible area is covered by ocean (Fig.\ref{fig:snapshot}).

\section{Effects of clouds on photometric characterization of Earth-like Exoplanets}
\label{s:dis}

Our primary goal in developing an improved treatment of forward and inverse procedures is to allow us to examine the effects of clouds. 

The presence of clouds affects the recoverable  fractional areas of
the different surface components in at least three ways: blocking the direct reflection spectra from the underlying surface areas, the nonlinear 
modification of the underlying surface reflection spectra by semi-transparent clouds, and the additional and strong time variability they introduce.
We examine these three effects in turn below.

\subsection{Cloud cover of the underlying planetary surface}

Approximately half of the surface of the Earth is covered by clouds with
$\tau _{\rm cld} \ge 1$. To the extent that the  corresponding regions are completely 
unseen, the total surface area that one can directly observe is reduced by the same amount. 
Moreover the relative fractions of surface components contributing the light curves are modified because ocean and vegetation surfaces are more likely to be cloud-covered than soil \citep[see also][]{cowan2009}. 

\begin{figure}[htbp]
  \centerline{\includegraphics[width=200mm]{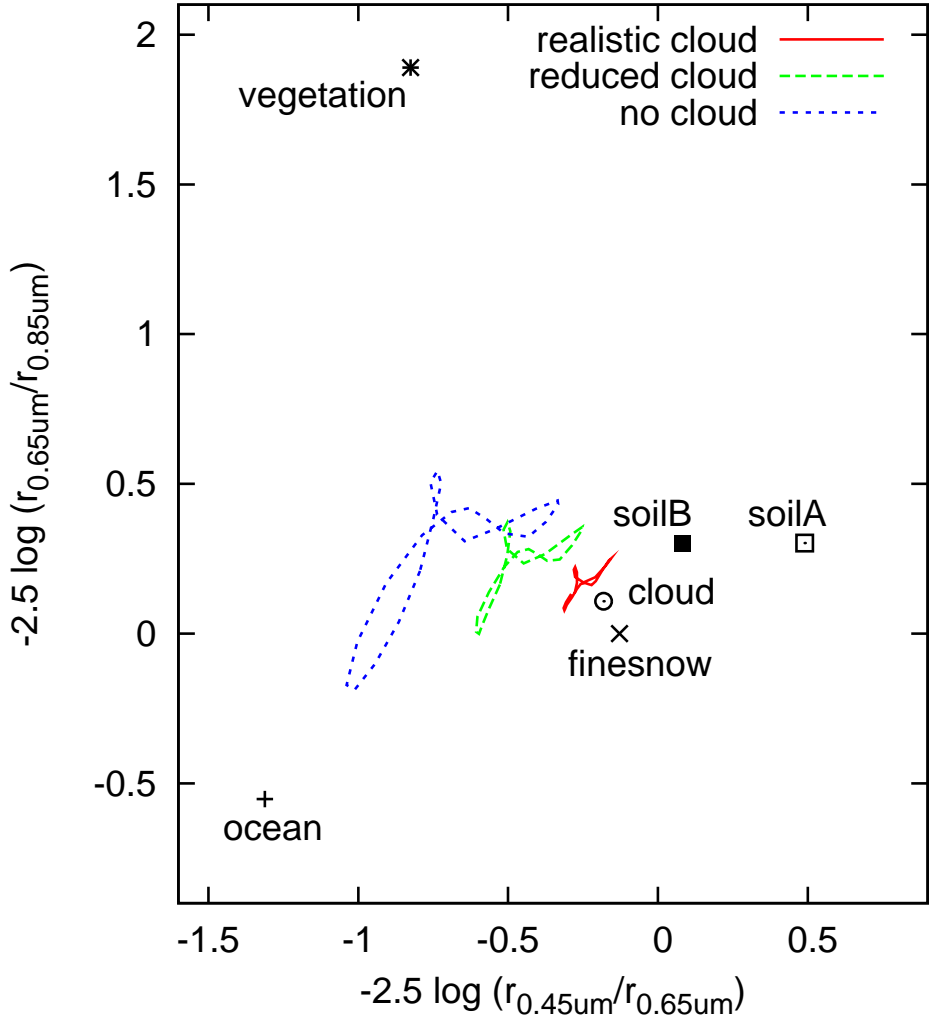}}
   \caption{Color-color diagram. The trajectories of the simulated light curve of our ``forward'' model with realistic cloud cover, that of the ``reduced cloud'' model, and that of ``no cloud model'' are displayed together along with the representative colors of various components indicated by points. }
  \label{fig:colorcolor_24-46}
\end{figure}

This effect of clouds may be evaluated more quantitatively with the
forward calculation.  To be specific, we consider hypothetical Earth-twins
with different cloud cover fractions; cloudless, reduced cloudiness, and realistic cloudiness cases.  The realistic cloud case is based on the MODIS 
cloud data as described in Section \ref{s:sim}, while the reduced cloudiness case has roughly 50 \% of the cloud cover fraction. 
The latter is created by multiplying both the observed cloud cover fraction and cloud optical thickness at each pixel by a uniformly distributed  random number between 0 and 1. 
The cloudless case neglects clouds entirely. 

The resultant color-color plot for $0.65\mu {\rm m}$ and $0.85\mu {\rm m}$ versus $0.45\mu {\rm m}$ and $0.65\mu {\rm m}$ is shown in Figure \ref{fig:colorcolor_24-46}.
The color trajectories over a diurnal period of the Earth are shown by the three curves.  For reference, the
colors of hypothetical planets with uniform surfaces of ocean (plus), soil A (open square), soil B (painted square), vegetation (asterisk), and snow (cross) with the US standard atmosphere are plotted as points. 
This figure indicates clearly why clouds dramatically limit our ability to recover the surface components from light curves. 
As the cloud cover fraction increases, 
the overall color approaches that of the clouds, and the amplitude of color variations is significantly reduced. 
The qualitative behavior is similar for color-color plots in other bands. 

It is, however, encouraging that the clouds do not much change the overall shape of the trajectory in color-color space. This implies that the recovery of surface components is still possible given sufficient photometric precision.

\subsection{Modification of the surface reflection spectra by semi-transparent cloud cover}
\label{ss:dis2}

Our inverse procedure assumes that the observed total albedo spectrum of the Earth can be approximated by a linear combination of five distinct spectra corresponding to clouds and the four surface components.  In reality, the reflection spectrum of a particular
surface component covered by clouds is not necessarily composed of a sum
of the clouds and the surface spectra.

\begin{figure}[htbp]
\begin{minipage}{0.5\hsize}
  \centerline{\includegraphics[width=80mm]{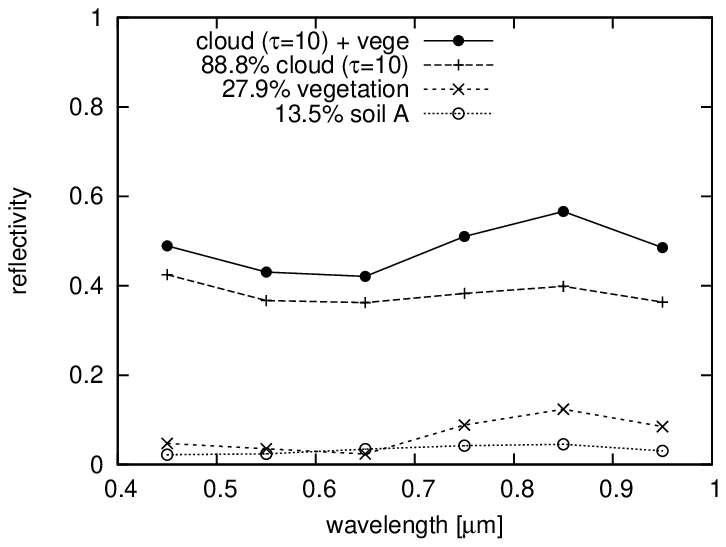}}
\end{minipage}	
\begin{minipage}{0.5\hsize}
  \centerline{\includegraphics[width=80mm]{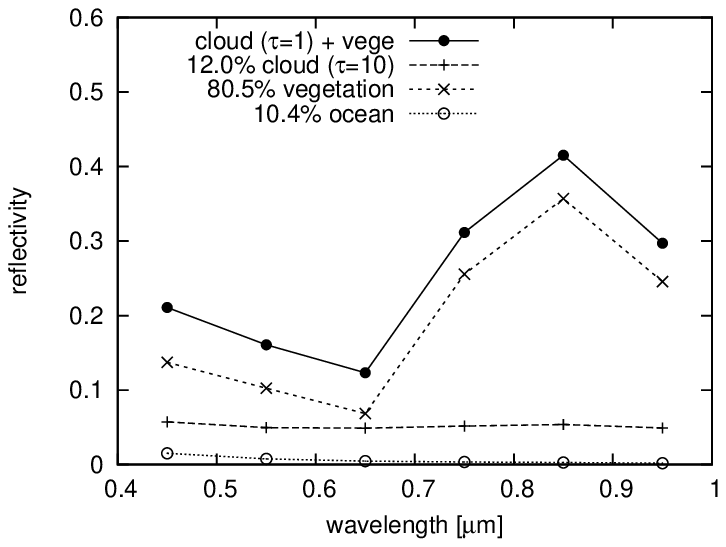}}
  \end{minipage}
  \caption{Spectra of vegetation as seen through 100 \% cloud cover and the resulting decomposition based on model I.  Left: the $\tau _{\rm cld}=10$ case. It is best-fit by a combination of 0 \% ocean, 0 \% snow, 27.9\% vegetation (cross), 13.5\% soil (open circle), 88.8\% cloud (plus).  Right: the $\tau _{\rm cld}=1$ case. It is best-fit by a combination of 10.4\% ocean (open circle), 0 \% snow, 80.5\% vegetation (cross) 12.0\%, 0 \% soil, and cloud (plus).  }
  \label{fig:cldml-10+grass}
\end{figure}

In order to show the difference more quantitatively, we use {\it rstar6b} to compute 
the reflectivity of vegetation entirely covered by clouds of $\tau_{\rm
cld} = 10$. As plotted in the left panel of Figure
\ref{fig:cldml-10+grass}, the resulting spectrum is best-fit by a sum
of clouds, vegetation and soil with fractional areas of 0.888, 0.279 and 0.135, respectively. 
The cloud-diluted red-edge feature of the vegetation is partially misidentified as soil in
this case. 
The estimated total area exceeds unity so as to compensate for the lower reflectivities of vegetation and soil compared to clouds. 
We repeat the exercise for vegetation covered by clouds of $\tau _{\rm cld} = 1$; the result is plotted in the right panel of Figure \ref{fig:cldml-10+grass}. In this case, the spectrum is decomposed as a sum of clouds, vegetation and ocean. 

These misidentifications are the result of our  simplified treatment of the template surface albedos assumed in the inverse procedure.  In reality, the wide range of cloud properties results in a variety of cloud spectra (Fig. \ref{fig:effalbd_cld}) that are partially
degenerate with the spectra of the other surface components. 
In principle, this issue could be addressed with a more elaborate and realistic model, but doing so is not justified without quite high signal-to-noise ratio observations.

\subsection{Additional time variability}

If an exoplanet is an oblique rotator like the Earth, the overall color of its surface may exhibit seasonal variations including those due to autumn leaves and snow cover.  Otherwise its surface components are taken to be independent of time. 
Clouds can introduce light-curve fluctuations due to their spatial and temporal variations, but the good agreement of the diurnal EPOXI light curves and our forward simulations based on a static cloud map (see Section \ref{s:sim}) indicates that the variability due to clouds is not important on a 24 hour time scale, as discussed in e.g. \citet{ford2001,palle2008}. 

Since we may have to stack the data of exoplanets over a period of a few weeks or more to improve the photon statistics, the cloud variations on
those longer time scales may well be more important \citep{ford2001,palle2008}. Nevertheless, we may hope that the decomposition of the stacked light curves will simply provide us an average measure of each component because the light curves are  essentially linear combinations of the observing days. 

In turn, the time variability among the light curves folded according to the diurnal period may act as a useful probe of cloud pattern and thus weather on an exoplanet. 
\citet{palle2008} found that the apparent rotation period is shifted by the net movement of cloud distribution. 
Also it might be possible to use the time variability to separate the contributions from clouds and those from surface components by e.g.   looking for the day with the lowest reflectivity because it is likely to correspond to the least cloudy day; generally surface components have lower reflectivity than clouds (Fig. \ref{fig:albedo_surface}). 

\section{Summary and discussion}
\label{s:conclusion}

We have presented a methodology to interpret future photometric data of 
Earth-like exoplanets. 
The methodology is an improvement over the one presented in Paper I in various ways. 
In particular we have incorporated the presence of clouds, utilized the radiative transfer code {\it  rstar6b} for the atmosphere and based the reconstruction on a more diverse set of template albedo spectra. 
In addition, we have validated the forward procedure  using the EPOXI observations of the Earth's light curves as well as using them as an input for the inverse procedure. 

Our main conclusions are summarized below.

\begin{enumerate}
\item Our forward procedure reproduces the EPOXI light curves well using the spatial distribution of the cloud cover fraction and optical depth on each observing day. 
This scheme will be useful to predict the light curves of Earth-like exoplanets, potentially detectable with future ground-based and space facilities. 
\item While our inverse procedure is approximate and model-dependent, the presence of ocean and clouds is routinely detectable from the diurnal light curves of an Earth-twin with S/N$\gtrsim $ 10. However, S/N $\gtrsim $ 20 is required for a high confidence detection. 
\item For the necessarily limited suite of models considered in Section \ref{ss:modeldepend}, the systematic uncertainties associated with the assumptions and input template spectra are similar in amplitude to the statistical errors at S/N$\sim $10. 
Thus, systematic uncertainties are not necessarily  dominant for the signal-to-noise levels realistically expected for early exoplanet light-curve data. 
However, it is obviously possible for a terrestrial planet to have important surface components without counterparts on the Earth; if so, very serious systematic errors could ensue.  
\item For an Earth-twin exoplanet, S/N $\gtrsim 100$ may enable us to detect the presence of components other than ocean and clouds in a fairly model-independent fashion.
\end{enumerate}

As discussed in Section \ref{s:dis}, cloud cover adds complexity and uncertainty to the interpretation of planetary light curves. 
Fortunately, clouds have their unique features such as the grey  reflection spectra, time variability \citep[e.g.][]{ford2001,palle2008}, the effect on thermal emission \citep[e.g.][]{tinetti2006a, tinetti2006b, kitzmann2011}, the anisotropic nature of scattering \citep[e.g.][]{robinson2010, dekok2011}, and polarization \citep[e.g.][]{karalidi2011}. 
In future work, it would be desirable to develop techniques for removal of clouds by making use of these characteristics in order to extract uncontaminated information on the surface. 

Our identification of surface components depends on the differences in their reflection spectra.  
For example, the presence of 
ocean is identified as a ``dark'' components in our inversion procedure.
Although there might be other ``dark'' components than ocean, this
approach will provide a useful check for the presence of oceans in 
combination with other indicators, such as ``ocean glint'' \citep{williams2008,oakley2009,robinson2010} or polarimetry \citep{zugger2010}.

In this paper as well as Paper I, we have focused on the case of an Earth-twin, but in order to further investigate the potential of photometric light curves for the characterization of terrestrial exoplanets, it will be necessary to consider models of planets unlike the Earth in various aspects. 
We plan to investigate such models in future work. 

\acknowledgments

We thank Atsushi Taruya for useful discussion and comments, Giovanna Tinetti for providing the reflection spectra shown in Figure 7 as a reference,
and Nicolas B. Cowan for his advice concerning the EPOXI data
processing. We are grateful to OpenCLASTR project for  providing the {\it rstar6b} package.  Y.F. thanks the staff of the Department o f Astrophysical Sciences of Princeton University  for their warm hospitality during her visit, in 
particular Dmitry Savransky, Tyler Groff, and David Spergel for their
helpful discussions.  Y.F. and Y.S. gratefully acknowledge support from
the Global Collaborative Research Fund (GCRF) ``A World-wide
Investigation of Other Worlds'' grant, and the Global Scholars Program
of Princeton University, respectively.  H.K. is supported by JSPS (Japan
Society for the Promotion of Science) Fellowship for Research
(PD:22-5467).  E.L.T is supported
in part by the GCRF grant and the World Premier International Research Center Initiative (WPI Initiative), MEXT, Japan.  This work is also supported by JSPS Core-to-Core Program ``International Research Network
for Dark Energy''.

{\appendix

\section{Pre-processing of the MODIS cloud data}
\label{ap:input}

The MODIS global cloud data available online\footnote{http://ladsweb.nascom.nasa.gov/} require  pre-processing to deal with two of their limitations: pixels with invalid cloud cover fraction data and those with invalid cloud optical thickness data. 

For the former case, we assign a {\it cloud cover fraction} equal to the average value of adjacent pixels with valid values, if any. 
We iterate this procedure to assign values to pixels which initially have no adjacent pixels with valid values. 
Re-binning to $2^{\circ}\times2^{\circ}$ resolution is accomplished by averaging. 

In the latter case, pixels may be invalid for one of three reasons: (1) the cloud optical thickness is too small to measure; these pixels are assigned a  value of 0. (2) the Sun illuminates regions near the poles at too low angle of incidence to allow measurements of cloud optical thickness; these pixels are assigned values via the same iterative averaging of adjacent pixels used for cloud cover fraction (see above). (3) 
A small fraction of the Earth was not observed by MODIS on the days of the EPOXI observations; the resulting gaps were filled by the same iterative procedure. 

Because the radiance is related to the cloud optical thickness in a non-linear way, it is inappropriate to average $\tau _{\rm cld}$ in the coarse-bining process. 
Instead, we assign the value of cloud optical thickness in one of the sub-pixels to the whole pixel.

\section{Longitutinal mapping of the surface}
\label{ap:aitoff}

Diurnal variations may be translated into a longitudinal map \citep{cowan2008} or even a 2-dimensional map if combined with yearly variations 
\citep{kawahara2010}. 
Here we map the diurnal variation of each component reconstructed from June data (The left panel of Figure \ref{fig:BVLS_June_cldml-10_soil-inceptisol}) into a 7-slice model \citep{cowan2008} and display the result in Figure \ref{fig:aitoff_June_cldml-10_soil-inceptisol}. 
It is apparent that some of the major geographical features of the Earth e.g. two oceans, the Sahara desert, and the two largest land masses are properly located. 

\begin{figure}[!h]
\begin{minipage}{0.33\hsize}
  \centerline{\includegraphics[width=50mm]{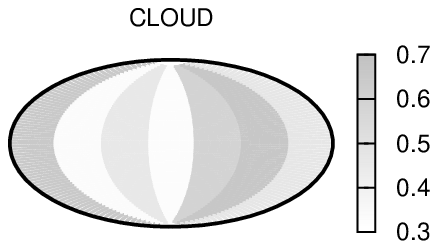}}
\end{minipage}
\begin{minipage}{0.33\hsize}
  \centerline{\includegraphics[width=50mm]{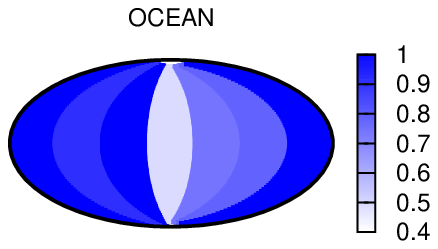}}
\end{minipage}
\begin{minipage}{0.33\hsize}
  \centerline{\includegraphics[width=50mm]{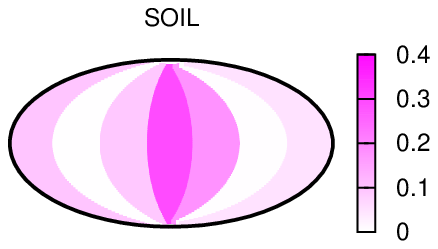}}
\end{minipage}
\begin{minipage}{0.33\hsize}
  \centerline{\includegraphics[width=50mm]{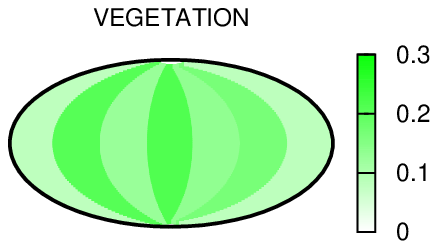}}
\end{minipage}
\begin{minipage}{0.33\hsize}
  \centerline{\includegraphics[width=50mm]{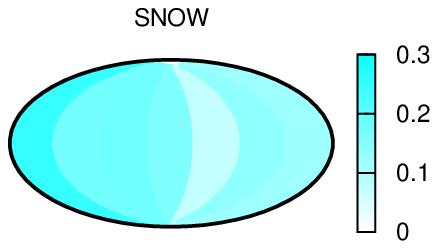}}
\end{minipage}
\begin{minipage}{0.33\hsize}
  \centerline{\includegraphics[width=50mm]{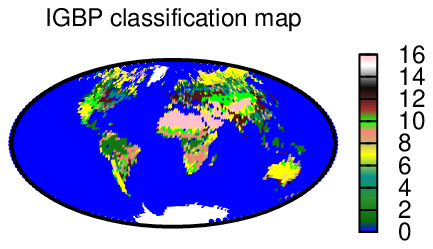}}
\end{minipage}
  \caption{The 7-slice longitudinal distribution of each component recovered from the June light curves. The indices of the IGBP classification map (the bottom left panel) are--- 0: ocean, 1: evergreen needleleaf forest, 2: evergreen broadleaf forest, 3: deciduous needleleaf forest, 4: deciduous broadleaf forest, 5: mixed forest, 6: closed shrubland, 7: open shrubland, 8: woody savannas, 9: savannas, 10: grasslands, 11: permanent wetlands, 12: croplands, 13: urban and built-up, 14: cropland/natual vegetation mosaic, 15: snow and ice, and 16: barren or sparsely vegetated (http://modis-atmos.gsfc.nasa.gov/ECOSYSTEM/index.html).}
  \label{fig:aitoff_June_cldml-10_soil-inceptisol}
\end{figure}

}

\end{document}